\documentclass[11pt]{article}

\title{Pairwise adaptive thermostats for improved accuracy and stability in dissipative particle dynamics}
\author{Benedict Leimkuhler and Xiaocheng Shang\footnote{Corresponding author. Current address: Division of Applied Mathematics, Brown University, 182 George Street, Providence, Rhode Island 02912, USA. Email: \href{mailto:x.shang@brown.edu}{x.shang@brown.edu} } \\
\small{School of Mathematics, University of Edinburgh, Edinburgh, EH9 3FD, UK} }
\date{\today}

\usepackage{cite}

\usepackage{hyperref}
\usepackage{graphicx}

\usepackage{caption}
\DeclareGraphicsExtensions{.eps,.mps,.pdf,.jpg,.png}
\graphicspath{{figures/}{../figures/}}

\usepackage{fancyhdr}

\usepackage{amsmath}
\usepackage{bm}
\usepackage{amsfonts}

\usepackage{dsfont}

\usepackage[titletoc,toc,title]{appendix}

\usepackage[margin=3.0cm]{geometry}% by courtesy of Mico

\newcommand{\Tr}{\mathrm{Tr}}

\renewcommand{\vec}[1]{{\mathbf #1}}

\newcommand{\q}{{\vec{q}}}
\newcommand{\p}{{\vec{p}}}

\newcommand{\M}{{\vec{M}}}
\newcommand{\I}{{\vec{I}}}

\newcommand{\dd}{{\rm d}}

\usepackage{mathrsfs}

\usepackage{soul}
\usepackage{color}

\usepackage{algorithm}
\usepackage{algorithmic}

\usepackage{url}
\usepackage{subfigure}

\usepackage{epstopdf}
\DeclareGraphicsExtensions{.eps,.mps,.pdf,.jpg,.png}

\usepackage{latexsym}
\usepackage{mathrsfs}
\usepackage{amssymb}
\usepackage{enumitem}

\usepackage{url}

\usepackage{array}
\newcolumntype{C}[1]{>{\centering\let\newline\\\arraybackslash\hspace{0pt}}m{#1}}

\begin{document}

\maketitle

\begin{abstract}
We examine the formulation and numerical treatment of dissipative particle dynamics (DPD) and momentum-conserving molecular dynamics. We show that it is possible to improve both the accuracy and the stability of DPD by employing a pairwise adaptive Langevin thermostat that precisely matches the dynamical characteristics of DPD simulations (e.g., autocorrelation functions) while automatically correcting thermodynamic averages using a negative feedback loop. In the low friction regime, it is possible to replace DPD by a simpler momentum-conserving variant of the Nos\'{e}--Hoover--Langevin method based on thermostatting only pairwise interactions; we show that this method has an extra order of accuracy for an important class of observables (a superconvergence result), while also allowing larger timesteps than alternatives. All the methods mentioned in the article are easily implemented. Numerical experiments are performed in both equilibrium and nonequilibrium settings; using Lees--Edwards boundary conditions to induce shear flow.
\end{abstract}

\pagenumbering{arabic}

\section{Introduction}

Classical molecular dynamics (MD), where the motion of individual atoms is governed by Newton's law in the microcanonical ensemble (where energy, i.e., the Hamiltonian of the system, is conserved), has been widely used in molecular simulations~\cite{Allen1989,Frenkel2001}. However, the constant energy setting is not relevant to a real-world laboratory environment since energy, as an extensive variable, depends on system size. In typical cases, one replaces the microcanonical ensemble by the canonical one, where temperature is conserved using suitable ``thermostat'' techniques.

One popular thermostat is Langevin dynamics, whereby each particle is subject to dissipative and collisional interactions with the particles of an artificial ``heat bath'' and modeled by supplementing the conservative Newtonian equations of motion with balanced damping and stochastic terms in such a way that the desired target system temperature is maintained. However, as pointed in~\cite{Hoogerbrugge1992}, in order to be consistent with hydrodynamics, a particle model should respect Galilean invariance, and, in particular, should conserve momentum, something that Langevin dynamics fails to do.   Fundamentally, Langevin dynamics and \mbox{its} overdamped Brownian dynamics limit, are appropriate for modeling systems in or \mbox{near} thermodynamic equilibrium and therefore do not take into account the possibility of an underlying fluid flow, thereby precluding their use in situations where the flow of the soft matter system cannot be predicted beforehand (e.g., when dealing with interfaces or nonuniform flow). Moreover, it has been reported in~\cite{Duenweg1993} that, due to the violation of global momentum conservation, Langevin dynamics can lead to nonphysical screening of hydrodynamic interactions with a screening length proportional to the inverse square root of the friction coefficient of the algorithm.  In order to control simulation artifacts, one is led to use a very small friction coefficient, effectively reducing Langevin dynamics to Hamiltonian dynamics in the microcanonical ensemble.   Therefore, when hydrodynamics is of interest, standard thermostats should be replaced by momentum-conserving thermostats, in particular the so-called dissipative particle dynamics (DPD) method of Hoogerbrugge and Koelman~\cite{Hoogerbrugge1992}.

DPD was first proposed in order to recover the properties of isotropy (and Galilean invariance) that were broken in the so-called lattice-gas automata method~\cite{Frisch1986}. In DPD, each body is regarded as a coarse-grained particle.   These particles interact in a soft (and short-ranged) potential, allowing larger integration timesteps than would be possible in MD, while simultaneously decreasing the number of degrees of freedom required. As in Langevin dynamics, a thermostat consisting of well-balanced damping and stochastic terms is applied to each particle. However, unlike in Langevin dynamics, both terms are pairwise and the damping term is based on relative velocities, leading to the conservation of both the angular momentum and the linear momentum. The property of Galilean invariance (i.e., the dependence on the relative velocity) makes DPD a profile-unbiased thermostat (PUT)~\cite{Evans1986,Evans2008} by construction and thus it is an ideal thermostat for nonequilibrium molecular dynamics (NEMD)~\cite{Soddemann2003}. The momentum is expected to propagate locally (while global momentum is conserved) and thus the correct hydrodynamics is expected in DPD~\cite{Soddemann2003}, as demonstrated previously in~\cite{Espanol1995a}. Due to the aforementioned properties, DPD has been widely used to recover thermodynamic, dynamical, and rheological properties of complex fluids, with applications in polymer solutions~\cite{Symeonidis2005}, colloidal suspensions~\cite{Pan2010}, multiphase flows~\cite{Pan2014}, and biological systems~\cite{Li2013}.   DPD has been  compared with Langevin dynamics for out-of-equilibrium simulations of polymeric systems in~\cite{Pastorino2007}, where as expected the correct dynamic fluctuations of the polymers were obtained with the former but not with the latter.

Given its promising prospects from the applications perspective, and its widespread use in large scale simulations,  the optimal design of numerical methods for DPD becomes crucially important, in particular the numerical efficiency in practice~\cite{Besold2000,Vattulainen2002,Nikunen2003,Chaudhri2010}.  Numerous numerical schemes~\cite{Besold2000,Lowe1999,DeFabritiis2006,Shardlow2003,Peters2004,Stoyanov2005,Allen2007} have been proposed in the last two decades following the introduction of DPD, which are intended to reduce nonphysical artifacts (especially in the large stepsize regime) induced by the discretization error. Recently, we have systematically examined the performance (in terms of accuracy, efficiency, and robustness) of a number of the most popular methods in the literature~\cite{Leimkuhler2015}.

In addition, we have proposed in~\cite{Leimkuhler2015} an alternative stochastic momentum-conserving thermostat, the pairwise Nos\'{e}--Hoover--Langevin (PNHL) thermostat.   This method mimics the DPD system in the regime of low friction, however achieving much higher accuracy and computational efficiency.   One contribution of the current article is a perturbation analysis showing that averages of observables of a certain (common) form performed using a nonsymmetric splitting of the PNHL system (i.e., the PNHL-N method~\cite{Leimkuhler2015}) have unexpected second order accuracy (as a power of the stepsize), justifying the enhanced performance of PNHL observed in simulations.

The second important contribution of this article is a new pairwise adaptive Langevin (PAdL) thermostat to replace DPD in the regime of moderate or high friction.  This method draws on work on adaptive thermostats~\cite{Jones2011,Leimkuhler2015a,Shang2015}, by supplementing a DPD type pairwise stochastic perturbation by an auxiliary control law (also pairwise) to maintain the thermodynamic state.   The new method fully captures the dynamics of DPD (for example, autocorrelation functions in DPD are precisely reproduced) and thus can be directly applied in the setting of momentum-conserving simulations as a replacement for DPD.  We describe a simple splitting-based numerical method for PAdL.  While PAdL has similar per-timestep computational cost, the method is shown to generate substantially more accurate approximations to thermodynamic averages at the same stepsize as DPD (as much as an order of magnitude). Moreover, and perhaps more significantly, the stepsize can be increased by around 50\% using PAdL, for similar accuracy, resulting in a much more efficient overall simulation method.

Furthermore, we discuss the proper treatment of Lees--Edwards boundary conditions in the DPD setting, an essential part of modeling shear flow.

The rest of the article is organized as follows.  In Section~\ref{sec:DPD}, we review the formulation of DPD and the momentum-conserving PNHL method, and introduce the newly proposed PAdL thermostat that mimics the dynamics of DPD.  We investigate in Section~\ref{sec:Numerical_Methods} numerical methods for PNHL and PAdL and give results on the order of accuracy of various schemes, in particular showing that the PNHL-N method is second order in its approximation of ergodic (long time) averages of a certain class of observables. Section~\ref{sec:Numerical_Experiments} presents  numerical experiments in both equilibrium and nonequilibrium cases, comparing the performance of numerous popular numerical methods in practice. The proper treatment of Lees--Edwards boundary conditions in the context of momentum-conserving thermostats is also discussed in Section~\ref{sec:Numerical_Experiments}. Our findings are summarized in Section~\ref{sec:Conclusions}.

\section{Dissipative particle dynamics and pairwise thermostats}
\label{sec:DPD}

In this section, we review the formulation of DPD and the momentum-conserving PNHL thermostat, followed by the introduction of the PAdL thermostat.

\subsection{Dissipative particle dynamics (DPD)}

The original DPD system was updated in discrete time steps and was later reformulated by Espa\~{n}ol and Warren~\cite{Espanol1995} as a system of It\={o} stochastic differential equations (SDEs).

We write the DPD system in a compact (vector) form:
\begin{equation}
  \label{eq:DPD}
  \begin{aligned}
    \dd \mathbf{q} &= \mathbf{M}^{-1}\mathbf{p} \dd t \, , \\
    \dd \mathbf{p} &= -\nabla U(\mathbf{q})\dd t - \gamma\boldsymbol{\Gamma}(\mathbf{q}) \mathbf{M}^{-1} \mathbf{p} \dd t + \sigma \boldsymbol{\Sigma}(\mathbf{q}) \dd \mathbf{W} \, ,
  \end{aligned}
\end{equation}
where $\mathbf{q}$ and $\mathbf{p}$ are $dN$-dimensional vectors, $d$ being the underlying dimensionality of the physical space (typically $d=3$) and $N$ being the number of particles, representing positions and momenta of particles, respectively, $\mathbf{M}$ is the diagonal mass matrix, $-\nabla U$ is the conservative force given in terms of a potential energy function $U=U(\mathbf{q})$, $\gamma$ is the friction coefficient and $\sigma$ represents the strength of the random forces, $\mathbf{W}$ is a vector of $S=dN(N-1)/2$ independent Wiener processes (note that, starting from the first element, every $d$ consecutive elements are identical, intended for each interacting pair, and the symmetry of $\dd \mathrm{W}_{ij} = \dd \mathrm{W}_{ji}$ is required to ensure the momentum conservation), and the  matrices $\boldsymbol{\Gamma}(\mathbf{q}) \in \mathbb{R}^{dN \times dN}$ and $\boldsymbol{\Sigma}(\mathbf{q}) \in \mathbb{R}^{dN \times S}$ satisfy the following relation:
\begin{equation} \label{flucdiss}
  \boldsymbol{\Gamma} = \boldsymbol{\Sigma} \boldsymbol{\Sigma}^{T} \, ,
\end{equation}
which can be thought of as a generalized fluctuation-dissipation relation.

The matrix $\boldsymbol{\Gamma}(\mathbf{q})$ may be written explicitly as
\begin{equation}\label{eq:matrix_Gamma}
\boldsymbol{\Gamma}(\mathbf{q}) =
\left(
  \begin{array}{cccc}
    \displaystyle\sum_{j\neq 1}\omega^{\mathrm{D}}_{1j}\mathbf{E}_{1j} & -\omega^{\mathrm{D}}_{12}\mathbf{E}_{12} & \cdots & -\omega^{\mathrm{D}}_{1N}\mathbf{E}_{1N} \\
    -\omega^{\mathrm{D}}_{21}\mathbf{E}_{21} & \displaystyle\sum_{j\neq 2}\omega^{\mathrm{D}}_{2j}\mathbf{E}_{2j} & \cdots & -\omega^{\mathrm{D}}_{2N}\mathbf{E}_{2N} \\
    \vdots & \vdots & \ddots & \vdots \\
    -\omega^{\mathrm{D}}_{N1}\mathbf{E}_{N1} & -\omega^{\mathrm{D}}_{N2}\mathbf{E}_{N2} & \cdots & \displaystyle\sum_{j\neq N}\omega^{\mathrm{D}}_{Nj}\mathbf{E}_{Nj} \\
  \end{array}
\right) \, ,
\end{equation}
where $\omega^{\mathrm{D}}_{ij} = \left[ \omega^{\mathrm{R}}_{ij} \right]^{2}$ is the weight function defined in the DPD system and $\mathbf{E}_{ij}=\hat{\mathbf{q}}_{ij}\hat{\mathbf{q}}_{ij}^{T}$, where $\hat{\mathbf{q}}_{ij} = (\mathbf{q}_{i} - \mathbf{q}_{j})/r_{ij}$ is the unit vector with $r_{ij} = \| \mathbf{q}_i - \mathbf{q}_j\|$ being the distance between two particles, is the $d$ by $d$ projection matrix on particles $i$ and $j$.

In the standard treatment of DPD, the conservative forces are derived from a sum of pair potentials
\begin{equation}
U(\mathbf{q}) = \sum_{i=1}^{N-1} \sum_{j=i+1}^N \varphi(r_{ij}) \, ,
\end{equation}
a typical choice of $\varphi$ being~\cite{Groot1997}
\begin{equation}\label{eq:Conservative_Force}
\varphi(r_{ij})=
  \begin{cases}
    \displaystyle \frac{a_{ij} r_{\mathrm{c}}}{2} \left(1-\frac{r_{ij}}{r_{\mathrm{c}}}\right)^{2} \, , &r_{ij}<r_{\mathrm{c}} \, ;\\
    \quad \quad \quad 0 \, , & r_{ij}\geq r_{\mathrm{c}} \, ,
  \end{cases}
\end{equation}
where parameters $a_{ij}$ represent the maximum repulsion strengths between interacting pairs, and $r_{\mathrm{c}}$ denotes a cutoff radius that is used in order to reduce the computational cost by restricting the number of pairs that need to be involved in the force computation.

In more recent studies, the conservative force is frequently obtained by coarse-graining procedures (e.g.,~\cite{Lyubartsev1995,Lei2010,Li2014a}) and may be represented by tabled data or an interpolation thereof~\cite{Suter2014}.

The weight function can be arbitrarily chosen without violating the thermal equilibrium. A simple, popular choice is
\begin{equation}\label{eq:Weight_Function_R}
  \omega^{\mathrm{R}}_{ij} = \omega^{\mathrm{R}}(r_{ij})=
  \begin{cases}
  \displaystyle 1-\frac{r_{ij}}{r_{\mathrm{c}}} \, , & r_{ij}<r_{\mathrm{c}} \, ;\\
  \quad 0 \, , & r_{ij}\geq r_{\mathrm{c}} \, .
  \end{cases}
\end{equation}

\subsection{Statistical properties of the DPD system in equilibrium}

Let $H$ represent the system Hamiltonian
\begin{equation}
  H(\mathbf{q},\mathbf{p}) = \displaystyle \frac{{\p}^T\M^{-1}{\p}}{2} + U(\mathbf{q})
\end{equation}
and assume the following fluctuation-dissipation relation:
\begin{equation}\label{eq:FDT}
  \sigma^{2}=2\gamma k_{\mathrm{B}}T \, ,
\end{equation}
where $k_{\mathrm{B}}$ is the Boltzmann constant and $T$ the temperature, it is easy to show that the DPD system~\eqref{eq:DPD} preserves the momentum-constrained canonical ensemble with density
\begin{equation}\label{eq:Gibbs_DPD}
  \rho_{\beta}(\mathbf{q},\mathbf{p}) = {Z}^{-1} \exp(-\beta H(\mathbf{q},\mathbf{p})) \times \delta\!\left[ \sum_i p^{x}_{i}- \pi_x \right] \delta\!\left [\sum_i p^{y}_{i} - \pi_y \right] \delta\!\left [\sum_i p^{z}_{i} - \pi_z \right] \, ,
\end{equation}
where $Z$ is a suitable normalizing constant (the partition function), $\beta^{-1}=k_{\mathrm{B}}T$, and $\boldsymbol{\pi}=(\pi_x,\pi_y,\pi_z)$ is the linear momentum vector. Additional constraints should be included  if the angular momentum is also conserved.  In an open system model, DPD conserves both angular and linear momenta (due to the fact that the interactions between particles depend on relative velocities) thus DPD is an isotropic Galilean-invariant thermostat which also preserves hydrodynamics~\cite{Soddemann2003,Allen2007,Moeendarbary2009}. However, it should be pointed out that if periodic boundary conditions are used (even along one direction), the conservation of angular momentum is lost, which is also expected when Lees--Edwards boundary conditions are applied.

Due to the use of coarse-grained potentials, and depending on the choice of weight functions, the ergodicity of stochastic thermostats for mesoscale modeling cannot be taken for granted. The ergodicity of the DPD system has been demonstrated only in the case of high particle density in one dimension by Shardlow and Yan~\cite{Shardlow2006}. It is also worth noting that the simulation study of Pastorino et al.~\cite{Pastorino2007} appears to contradict ergodicity when (i) soft DPD potentials are used and (ii) the number of interactions is limited sufficiently.

\subsection{Pairwise Nos\'{e}--Hoover--Langevin (PNHL) thermostat}

We have recently proposed the PNHL thermostat~\cite{Leimkuhler2015} in an attempt to improve the accuracy and stability in DPD simulations. Comparing to the standard formulation of DPD, in PNHL, the stochastic term was completely removed and the constant friction coefficient in the damping term was replaced by an additional dynamical variable that was driven by the difference between the instantaneous temperature based on relative velocities and the target temperature. Moreover, a Langevin thermostat was applied to the auxiliary variable in order to enhance the ergodicity. As in DPD, the PNHL thermostat conserves the momentum and is Galilean-invariant, thus correct hydrodynamics is expected to be preserved.

The equations of motion of the PNHL thermostat are given by
\begin{equation}
  \label{eq:PNHL}
  \begin{aligned}
    \dd \mathbf{q} &= \mathbf{M}^{-1} \mathbf{p}\dd t \, , \\
    \dd \mathbf{p} &= -\nabla U(\mathbf{q})\dd t - \xi \boldsymbol{\Gamma}(\mathbf{q})\mathbf{M}^{-1}\mathbf{p}\dd t \, , \\
    \dd \xi        &= G(\mathbf{q},\mathbf{p})\dd t - \tilde{\gamma} \xi \dd t + \tilde{\sigma} \dd \mathrm{W} \, ,
  \end{aligned}
\end{equation}
where $\xi$ is an auxiliary dynamical variable and $G(\mathbf{q},\mathbf{p})$ denotes the accumulated deviation of the instantaneous temperature away from the target temperature
\begin{equation}
  \label{eq:PNHL_G}
    G(\mathbf{q},\mathbf{p}) = {\mu}^{-1} \left[ \left(\mathbf{M}^{-1}\mathbf{p}\right)^{T} \boldsymbol{\Gamma}(\mathbf{q})\left(\mathbf{M}^{-1}\mathbf{p}\right) - \beta^{-1} \Tr \left(\boldsymbol{\Gamma}(\mathbf{q})\mathbf{M}^{-1}\right) \right] \, ,
\end{equation}
where $\mu$ is a coupling parameter which is referred to as the ``thermal mass'', coefficient constants $\tilde{\gamma}$ and $\tilde{\sigma}$ satisfy the following fluctuation-dissipation relation:
\begin{equation}
  {\tilde{\sigma}}^{2}=2\tilde{\gamma} k_{\mathrm{B}}T/\mu \, ,
\end{equation}
and $\mathrm{W}=\mathrm{W}(t)$ is a standard Wiener process.

The special case $\tilde{\gamma}=\tilde{\sigma}=0$ of the PNHL thermostat reduces the system to the  pairwise Nos\'{e}--Hoover (PNH) thermostat by Allen and Schmid~\cite{Allen2007}.  However, as mentioned above, ergodicity is not expected in some coarse-grained models, even with the addition of stochastic forces, thus, in particular, the PNH thermostat is likely to fail for many choices of weight functions and potentials.   The inclusion of noise through the auxiliary variable has been rigorously shown to restore ergodicity to the system, albeit only in the case of a nonpairwise scheme~\cite{Leimkuhler2009}.

The PNHL thermostat~\eqref{eq:PNHL} preserves the canonical ensemble with a modified invariant distribution (comparing to $\rho_{\beta}$~\eqref{eq:Gibbs_DPD}) in the standard DPD system
\begin{equation}\label{eq:Gibbs PNHL}
  \begin{aligned}
  \tilde{\rho}_{\beta}(\mathbf{q},\mathbf{p},\xi) = \, & Z^{-1} \exp\left({-\beta H(\mathbf{q},\mathbf{p})}\right) \exp\left({-\beta\mu\xi^{2}/2}\right) \\
  & \,  \times \delta\!\left[ \sum_i p^{x}_{i}- \pi_x \right] \delta\!\left [\sum_i p^{y}_{i} - \pi_y \right] \delta\!\left [\sum_i p^{z}_{i} - \pi_z \right]  \, .
  \end{aligned}
\end{equation}
We conjecture that the PNHL system is ergodic for this distribution if the weight functions have a sufficiently large support.

Two different splitting methods have been proposed in~\cite{Leimkuhler2015} for the PNHL thermostat, with the first being of a symmetric manner, labeled as PNHL-S, and the other nonsymmetric, PNHL-N (see details in Appendix~\ref{sec:Appendix_Schemes}). Both PNHL methods have been compared to various popular schemes for a number of physical quantities in~\cite{Leimkuhler2015}, and it turns out that both methods (especially the PNHL-N method) achieve significant improvements in terms of accuracy, robustness, and numerical efficiency over alternatives.    Both mathematical theory and numerical experiment with methods for Langevin dynamics (see, e.g.,~\cite{Leimkuhler2013c}) have repeatedly shown the efficiency advantage of symmetric numerical methods, thus the numerical results reported in~\cite{Leimkuhler2015} were, until now, a curious anomaly.  In the next section of this article, we finally demonstrate that PNHL-N has a superconvergence property (an extra order of accuracy) for averages of an important class of observables, thus explaining its numerical performance in practice.

The dynamical properties of the PNHL formulation correspond to those of the standard DPD system only in the low friction regime. In what follows, we propose a new momentum-conserving thermostat in order to have full control of the dynamics.

\subsection{Thermodynamically-corrected and hydrodynamics-preserving pairwise adaptive Langevin (PAdL) thermostat}
\label{sec:PAdL}

In order to maximize numerical efficiency (especially for large scale simulations), a large timestep is always preferred to discretize the system of interest. However, as already mentioned, a large timestep can result in pronounced nonphysical artifacts.  A recent article~\cite{Sivak2013} interprets the effect of the discretization error in Langevin dynamics as a means of driving the system away from a desired invariant distribution---excess energy is pumped into the system in the form of ``shadow work'' to prevent it from maintaining the desired temperature. Furthermore, if the forces are computed ``on-the-fly'' (see, e.g.,~\cite{Li2015b}), they are likely subject to substantial errors, which would also effectively ``heat'' the system up (see more discussions in~\cite{Leimkuhler2015a}).   As an illustration,  the hybrid quantum mechanics/molecular mechanics (QM/MM) method introduces localized heating due to the force-mixing at the boundary of the coupled QM and MM regions~\cite{Mones2014}. Moreover, the use of external fields in nonequilibrium models causes viscous heating (i.e., the energy pumped into the system leads to a temperature rise under steady perturbation), thus, in those cases, proper thermostats are required to drain the excess energy in order to maintain the correct system temperature~\cite{Backer2005}.

One way to regulate excess heat in mechanical systems is via the use of negative feedback loop controls.   Nos\'{e}--Hoover~\cite{Nose1984a,Hoover1991} is one such feedback control system, but the observation of~\cite{Jones2011} is that feedback loop controls can be introduced in tandem with a Langevin thermostat (their so-called ``adaptive'' Langevin dynamics), benefiting from the strong ergodicity properties of Langevin dynamics together with an automatic regulation of the kinetic energy.   These methods were further explored in~\cite{Leimkuhler2015a,Shang2015}.  Notably, in~\cite{Leimkuhler2015a} it was shown that an adaptive Langevin device could be used to dissipate excess heat due to noisy forces, while providing statistical convergence properties very similar to those of Langevin dynamics.

It is natural to consider extending the adaptive thermostat idea to DPD, thus hopefully providing correction for thermodynamic observables while mimicking, to a large extent, the properties of DPD. To this end,  we propose here a momentum-conserving PAdL thermostat, whose equations of motion are given by
\begin{equation}
  \label{eq:PAdL}
  \begin{aligned}
    \dd \mathbf{q} &= \mathbf{M}^{-1} \mathbf{p}\dd t \, , \\
    \dd \mathbf{p} &= -\nabla U(\mathbf{q})\dd t - \xi \boldsymbol{\Gamma}(\mathbf{q})\mathbf{M}^{-1}\mathbf{p}\dd t + \sigma \boldsymbol{\Sigma}(\mathbf{q}) \dd \mathbf{W} \, , \\
    \dd \xi        &= G(\mathbf{q},\mathbf{p})\dd t \, ,
  \end{aligned}
\end{equation}
where $\sigma$ is a constant coefficient as in DPD and the function $G$ is given in~\eqref{eq:PNHL_G}.   An additional Langevin thermostat could also be added on the auxiliary variable $\xi$ as in PNHL, but it is not necessary here because of the presence of the additional stochastic term directly contacting the physical variables.

It can be demonstrated, modifying the argument of~\cite{Jones2011}, that the PAdL system preserves the canonical ensemble with a modified invariant distribution
\begin{equation}\label{eq:Gibbs_PAdL}
  \begin{aligned}
  \tilde{\rho}_{\beta}(\mathbf{q},\mathbf{p},\xi) = \, & Z^{-1} \exp\left({-\beta H(\mathbf{q},\mathbf{p})}\right) \exp\left({-\frac{\beta\mu}{2}(\xi-\hat{\gamma})^{2}}\right) \\
  & \,  \times \delta\!\left[ \sum_i p^{x}_{i}- \pi_x \right] \delta\!\left [\sum_i p^{y}_{i} - \pi_y \right] \delta\!\left [\sum_i p^{z}_{i} - \pi_z \right]  \, ,
  \end{aligned}
\end{equation}
where $\hat{\gamma}$ can be thought of as the ``effective friction'' and the following fluctuation-dissipation relation is satisfied as in DPD:
\begin{equation}
  {\sigma}^{2} = 2\hat{\gamma} k_{\mathrm{B}}T \, .
\end{equation}

The invariant distribution~\eqref{eq:Gibbs_PAdL} implies that the auxiliary variable $\xi$ is Gaussian distributed with mean $\hat{\gamma}$ and variance $\beta^{-1}\mu^{-1}$. The auxiliary variable will fluctuate around its mean value during simulation and we can tune the value of the effective friction in order to recover the dynamics of DPD in a wide range of friction regimes. Therefore, we can think of the PAdL thermostat as the standard DPD system with an adaptive friction coefficient. It can be seen that the PAdL thermostat inherits essential properties of DPD (such as Galilean invariance and momentum conservation) required for consistent hydrodynamics. In the large thermal mass limit (i.e., $\mu \rightarrow \infty$), the PAdL thermostat effectively reduces to the standard DPD formulation~\eqref{eq:DPD}.

\section{Numerical methods for pairwise thermostats}
\label{sec:Numerical_Methods}

A great deal of effort has been devoted to developing accurate and efficient numerical methods to solve DPD and related systems. We have compared a number of popular schemes in~\cite{Leimkuhler2015}. In what follows we briefly review the numerical methods for the PNHL thermostat that will be included for further investigation in this article, followed by the derivation of the numerical scheme for the newly proposed PAdL thermostat. We also discuss the accuracy of equilibrium averages for those methods, in particular, we show the unexpected second order convergence of the PNHL-N method for certain observables.

\subsection{Numerical Treatment of PNHL}

Since we are unable to solve the PNHL system~\eqref{eq:PNHL} ``exactly'', we instead split the vector field of the system into pieces (A, B, C, D, and O)
%\begin{equation}
%  \begin{aligned}
%  \dd \left[ \begin{array}{c} \q \\ \p \\ \xi \end{array} \right] = \underbrace{\left[ \begin{array}{c} {\M}^{-1}\p \\ \vec{0} \\ 0 \end{array} \right] \dd t}_\mathrm{A} + \underbrace{\left[ \begin{array}{c} \vec{0} \\  -\nabla U(\q) \\ 0 \end{array} \right] \dd t}_\mathrm{B} + \underbrace{\left[ \begin{array}{c} \vec{0} \\ - \xi \boldsymbol{\Gamma}(\q)\mathbf{M}^{-1}\p \\ 0 \end{array} \right] \dd t}_\mathrm{C} + \underbrace{\left[ \begin{array}{c} \vec{0} \\ \vec{0} \\ G(\q,\p) \end{array} \right] \dd t}_\mathrm{D} + \underbrace{\left[ \begin{array}{c} \vec{0} \\ \vec{0} \\ - \tilde{\gamma} \xi \dd t + \tilde{\sigma} \dd \mathrm{W} \end{array} \right]}_\mathrm{O} \, ,
%  \end{aligned}
%\end{equation}
\begin{equation}
  \begin{aligned}
  \dd \left[ \begin{array}{c} \q \\ \p \\ \xi \end{array} \right] = \, & \underbrace{\left[ \begin{array}{c} {\M}^{-1}\p \\ \vec{0} \\ 0 \end{array} \right] \dd t}_\mathrm{A} + \underbrace{\left[ \begin{array}{c} \vec{0} \\  -\nabla U(\q) \\ 0 \end{array} \right] \dd t}_\mathrm{B} + \underbrace{\left[ \begin{array}{c} \vec{0} \\ - \xi \boldsymbol{\Gamma}(\q)\mathbf{M}^{-1}\p \\ 0 \end{array} \right] \dd t}_\mathrm{C} \\ & \, + \underbrace{\left[ \begin{array}{c} \vec{0} \\ \vec{0} \\ G(\q,\p) \end{array} \right] \dd t}_\mathrm{D} + \underbrace{\left[ \begin{array}{c} \vec{0} \\ \vec{0} \\ - \tilde{\gamma} \xi \dd t + \tilde{\sigma} \dd \mathrm{W} \end{array} \right]}_\mathrm{O} \, ,
  \end{aligned}
\end{equation}
in such a way that we can solve each subsystem exactly (see more details in~\cite{Leimkuhler2015}). We can write down the generators associated with each piece, respectively, as
\begin{align}
    \mathcal{L}_\mathrm{A} &= \M^{-1}\p \cdot \nabla_{\q} \, , \label{eq:generator_A} \\
    \mathcal{L}_\mathrm{B} &= -\nabla U(\q) \cdot \nabla_{\p} \, , \label{eq:generator_B} \\
    \mathcal{L}_\mathrm{C} &= - \xi \boldsymbol{\Gamma}(\q) \mathbf{M}^{-1}\p \cdot \nabla_{\p} \, , \\
    \mathcal{L}_\mathrm{D} &= G(\q,\p) \frac{\partial}{\partial\xi} \, , \label{eq:generator_D} \\
    \mathcal{L}_\mathrm{O} &= - \tilde{\gamma}\xi\frac{\partial}{\partial\xi} + \frac{{\tilde{\sigma}}^{2}}{2}\frac{\partial^{2}}{\partial\xi^{2}} \, ,
\end{align}
the sum of which is the generator of the PNHL system:
\begin{equation}\label{eq:generator_PAdL}
  \mathcal{L}_\mathrm{PNHL} = \mathcal{L}_\mathrm{A} + \mathcal{L}_\mathrm{B} + \mathcal{L}_\mathrm{C} + \mathcal{L}_\mathrm{D} + \mathcal{L}_\mathrm{O} \, .
\end{equation}

The flow map (or phase space propagator) of the system is given by
\begin{equation}
  { \cal F}_{t} = e^{t \mathcal{L} } \, ,
\end{equation}
where the exponential map represents the solution operator. Various approximations of ${\cal F}_t$ can be obtained as products (taken in different arrangements) of exponentials of the splitting terms, however, it turns out in a number of studies~\cite{Leimkuhler2013,Leimkuhler2013a,Leimkuhler2015b,Leimkuhler2015,Leimkuhler2015a} that different splittings and/or combinations give dramatically different performance in practice. Two different splitting methods of PNHL have been proposed in~\cite{Leimkuhler2015}, termed PNHL-S and PNHL-N, respectively:
\begin{equation}\label{eq:PNHL-S}
  e^{h\hat{\mathcal{L}}_\mathrm{PNHL-S}} = e^{\frac{h}{2}\mathcal{L}_\text{A}} e^{\frac{h}{2}\mathcal{L}_\mathrm{B}} e^{\frac{h}{2}\mathcal{L}_\mathrm{C}} e^{\frac{h}{2}\mathcal{L}_\mathrm{D}} e^{h\mathcal{L}_\mathrm{O}} e^{\frac{h}{2}\mathcal{L}_\mathrm{D}} e^{\frac{h}{2}\mathcal{L}_\mathrm{C}} e^{\frac{h}{2}\mathcal{L}_\mathrm{B}} e^{\frac{h}{2}\mathcal{L}_\mathrm{A}} \, ,
\end{equation}
and
\begin{equation}\label{eq:PNHL-N}
  e^{h\hat{\mathcal{L}}_\mathrm{PNHL-N}} = e^{\frac{h}{2}\mathcal{L}_\mathrm{A}} e^{\frac{h}{2}\mathcal{L}_\mathrm{B}} e^{\frac{h}{2}\mathcal{L}_\mathrm{C}} e^{\frac{h}{2}\mathcal{L}_\mathrm{D}} e^{h\mathcal{L}_\mathrm{O}} e^{\frac{h}{2}\mathcal{L}_\mathrm{D}} e^{\frac{h}{2}\mathcal{L}_\mathrm{C}} e^{\frac{h}{2}\mathcal{L}_\mathrm{A}} e^{\frac{h}{2}\mathcal{L}_\mathrm{B}} \, .
\end{equation}
The detailed integration steps of both methods can be found in Appendix~\ref{sec:Appendix_Schemes}. It is important to note that the steplengths associated with various operations are uniform and span the interval $h$. Thus the O step in either of the two methods is taken with a steplength of $h$, while the other pieces with a steplength of $h/2$. Note also that these two integrators differ only from the order of integrating the last two pieces. However, an additional force calculation is required (using the updated positions) in the PNHL-N scheme just before updating the last B piece at the end of each integration step. The additional force calculation could be costly, however, it was found to be offset by a great increase in accuracy and usable steplength~\cite{Leimkuhler2015}.

\subsection{Numerical Treatment of PAdL}

As in PNHL schemes, we separate the vector field of the PAdL system~\eqref{eq:PAdL} into pieces, which we label as A, B, O, and D, respectively:
%\begin{equation}
%  \dd \left[ \begin{array}{c} \q \\ \p \\ \xi \end{array} \right] = \underbrace{\left[ \begin{array}{c} {\M}^{-1}\p \\ \vec{0} \\ 0 \end{array} \right] \dd t}_\mathrm{A} + \underbrace{\left[ \begin{array}{c} \vec{0} \\  -\nabla U(\q) \\ 0 \end{array} \right] \dd t}_\mathrm{B} + \underbrace{\left[ \begin{array}{c} \vec{0} \\ - \xi \boldsymbol{\Gamma}(\mathbf{q})\mathbf{M}^{-1}\mathbf{p}\dd t + \sigma \boldsymbol{\Sigma}(\mathbf{q}) \dd \mathbf{W} \\ 0 \end{array} \right] }_\mathrm{O} + \underbrace{\left[ \begin{array}{c} \vec{0} \\ \vec{0} \\ G(\q,\p) \end{array} \right] \dd t}_\mathrm{D} \, ,
%\end{equation}
%\begin{equation}
%  \dd \left[ \begin{array}{c} \q \\ \p \\ \xi \end{array} \right] = \underbrace{\left[ \begin{array}{c} {\M}^{-1}\p \\ \vec{0} \\ 0 \end{array} \right] \dd t}_\mathrm{A} + \underbrace{\left[ \begin{array}{c} \vec{0} \\  -\nabla U \\ 0 \end{array} \right] \dd t}_\mathrm{B} + \underbrace{\left[ \begin{array}{c} \vec{0} \\ - \xi \boldsymbol{\Gamma}\mathbf{M}^{-1}\mathbf{p}\dd t + \sigma \boldsymbol{\Sigma} \dd \mathbf{W} \\ 0 \end{array} \right] }_\mathrm{O} + \underbrace{\left[ \begin{array}{c} \vec{0} \\ \vec{0} \\ G \end{array} \right] \dd t}_\mathrm{D} \, ,
%\end{equation}
\begin{equation}
  \begin{aligned}
  \dd \left[ \begin{array}{c} \q \\ \p \\ \xi \end{array} \right] = \, & \underbrace{\left[ \begin{array}{c} {\M}^{-1}\p \\ \vec{0} \\ 0 \end{array} \right] \dd t}_\mathrm{A} + \underbrace{\left[ \begin{array}{c} \vec{0} \\  -\nabla U(\q) \\ 0 \end{array} \right] \dd t}_\mathrm{B} \\ & \, + \underbrace{\left[ \begin{array}{c} \vec{0} \\ - \xi \boldsymbol{\Gamma}(\q)\mathbf{M}^{-1}\mathbf{p}\dd t + \sigma \boldsymbol{\Sigma}(\q) \dd \mathbf{W} \\ 0 \end{array} \right] }_\mathrm{O} + \underbrace{\left[ \begin{array}{c} \vec{0} \\ \vec{0} \\ G(\q,\p) \end{array} \right] \dd t}_\mathrm{D} \, .
  \end{aligned}
\end{equation}
Note that the generators of pieces A, B, and D here are exactly the same as defined in PNHL (Eqs.~\eqref{eq:generator_A}, ~\eqref{eq:generator_B}, and~\eqref{eq:generator_D}, respectively), while the generator for the remaining piece is given by
\begin{equation}
  \mathcal{L}_\mathrm{O} = - \xi \boldsymbol{\Gamma} \mathbf{M}^{-1}\p \cdot \nabla_{\p} + \frac{\sigma^{2}}{2} \boldsymbol{\Sigma}\boldsymbol{\Sigma}^{T}:\nabla_{\p}^{2} \, ,
\end{equation}
where $:$ denotes the Frobenius product for matrices, i.e., $\mathcal{A}:\mathcal{B} = \Tr(\mathcal{A}^{T}\mathcal{B})$. Note that the generator O here is different from that of PNHL, although they both represent Ornstein--Uhlenbeck processes. Overall, the generator for the PAdL system can be written as
\begin{equation}\label{eq:generator_PAdL}
  \mathcal{L}_\mathrm{PAdL} = \mathcal{L}_\mathrm{A} + \mathcal{L}_\mathrm{B} + \mathcal{L}_\mathrm{O} + \mathcal{L}_\mathrm{D} \, .
\end{equation}

It is straightforward to note that pieces A, B, and D can be solved exactly as in PNHL schemes. In fact, each interacting pair in O is also exactly solvable (in the sense of distributional fidelity)~\cite{Serrano2006}. For each interacting pair, subtracting $\dd \mathbf{v}_{j}$, where $\mathbf{v}_{j}$ is the velocity of particle $j$ (with corresponding mass $m_{j}$), from $\dd \mathbf{v}_{i}$ and multiplying by $\hat{\mathbf{q}}_{ij}$ on both sides yields
\begin{equation}
    m_{ij}\dd v_{ij} = - \xi \omega^{\mathrm{D}}(r_{ij})v_{ij}\dd t + \sigma \omega^{\mathrm{R}}(r_{ij})\dd \mathrm{W}_{ij} \, ,
\end{equation}
where $m_{ij}=m_{i}m_{j}/(m_{i}+m_{j})$ is the ``reduced mass'' and $v_{ij}=\hat{\mathbf{q}}_{ij} \cdot \mathbf{v}_{ij}$. With $\xi$ being a positive constant, the above equation is a standard Ornstein--Uhlenbeck process with the exact (in the sense of distributions) solution~\cite{Kloeden1992}
\begin{equation}\label{eq:OU_exact_PAdL}
    v_{ij}(t) =  e^{-\tilde{\tau} t} v_{ij}(0) + \sigma \sqrt{ \frac{1-e^{-2\tilde{\tau} t}}{2\xi m_{ij}} }\mathrm{R}_{ij} \, ,
\end{equation}
where $\tilde{\tau}=\xi \omega^{\mathrm{D}} / m_{ij}$, $v_{ij}(0)$ are the initial relative velocities, and $\mathrm{R}_{ij}$ are normally distributed variables with zero mean and unit variance. As demonstrated in~\cite{Leimkuhler2015a}, the exact solution above~\eqref{eq:OU_exact_PAdL} is still valid for $\xi < 0$. When $\xi=0$, one can simply replace $(1-e^{-2\tilde{\tau} t})/(2\xi m_{ij})$ by its well-defined asymptotic limit, in which case~\eqref{eq:OU_exact_PAdL} becomes
\begin{equation}
  v_{ij}(t) = v_{ij}(0) + \sigma (\omega^{\mathrm{R}}/m_{ij}) \sqrt{h} \mathrm{R}_{ij} \, .
\end{equation}
Then the velocity increments can be obtained as
\begin{equation}
    \Delta v_{ij} = v_{ij}(t) - v_{ij}(0) \, ,
\end{equation}
and subsequently the corresponding momenta can be updated by
\begin{align}
  \mathbf{p}^{n+1}_{i} &= \mathbf{p}^{n}_{i} + m_{ij}\Delta v_{ij} \hat{\mathbf{q}}^{n}_{ij} \, , \\
  \mathbf{p}^{n+1}_{j} &= \mathbf{p}^{n}_{j} - m_{ij}\Delta v_{ij} \hat{\mathbf{q}}^{n}_{ij} \, ,
\end{align}
which defines the propagator $e^{h\mathcal{L}_{\mathrm{O}_{i,j}}}$ for each interacting pair.

Various splitting methods could be constructed, we propose in this article a symmetric PAdL method (alternatively, the ``ABODOBA'' method), whose propagator can be written as
\begin{equation}\label{eq:PAdL-S}
  e^{h\hat{\mathcal{L}}_\mathrm{PAdL}} = e^{\frac{h}{2}\mathcal{L}_\mathrm{A}} e^{\frac{h}{2}\mathcal{L}_\mathrm{B}} e^{\frac{h}{2}\mathcal{L}_\mathrm{O}} e^{h\mathcal{L}_\mathrm{D}} e^{\frac{h}{2}\mathcal{L}_\mathrm{O}} e^{\frac{h}{2}\mathcal{L}_\mathrm{B}} e^{\frac{h}{2}\mathcal{L}_\mathrm{A}} \, ,
\end{equation}
where the O part associated with interacting pairs is given by
\begin{equation}
  e^{ \frac{h}{2}\hat{\mathcal{L}}_\mathrm{O} } =
  e^{ \frac{h}{2}\mathcal{L}_{\mathrm{O}_{N-1,N}} } \cdots
  e^{ \frac{h}{2}\mathcal{L}_{\mathrm{O}_{1,3}} }
  e^{ \frac{h}{2}\mathcal{L}_{\mathrm{O}_{1,2}} } \, .
\end{equation}
The detailed integration steps can be found in Appendix~\ref{sec:Appendix_Schemes}.

\subsection{Accuracy of equilibrium averages}
\label{sec:Order_of_Convergence}

We have demonstrated in~\cite{Leimkuhler2015} that a symmetric splitting method gives at least second order convergence to the invariant measure, and similar observations would hold for the symmetric methods described above, PNHL-S~\eqref{eq:PNHL-S} and the PAdL scheme~\eqref{eq:PAdL-S}.   Superconvergence properties (i.e., fourth order convergence) have also been proven in both Langevin dynamics~\cite{Leimkuhler2013} and adaptive Langevin thermostat~\cite{Leimkuhler2015a} for configurational sampling. Nonsymmetric splitting methods of geometric Langevin algorithms~\cite{Bou-Rabee2010} type can exhibit high order ergodic approximations, but first order convergence is generally expected for other nonsymmetric splitting methods.   This raises the obvious question of why we observed second order accuracy using the PNHL-N method~\eqref{eq:PNHL-N} in simulations performed in~\cite{Leimkuhler2015}.

The framework of long-time Talay--Tubaro expansion has been widely used in analyzing the accuracy of ergodic averages (i.e., averages with respect to the invariant measure) in stochastic numerical methods~\cite{Talay1990,Debussche2012,Leimkuhler2013,Leimkuhler2013a,Abdulle2014a,Abdulle2014,Leimkuhler2013c,Leimkuhler2015a,Leimkuhler2015b}. Therefore, in this section, we adopt the procedures to verify the second order convergence of the PNHL-N method for certain observables.

For a splitting method described by $ \mathcal{L}= \mathcal{L}_{\alpha} + \mathcal{L}_{\beta} + \dots + \mathcal{L}_{\zeta} $, its associated effective operator $\hat{\mathcal{L}}^{\dag}$ with stepsize $h$ is given by
\begin{equation}
  \exp\left(h \hat{\mathcal{L}}^{\dag}\right) = \exp\left(h \mathcal{L}_{\alpha}^{\dag}\right) \exp\left(h\mathcal{L}_{\beta}^{\dag}\right) \dots \exp\left(h\mathcal{L}_{\zeta}^{\dag}\right) \, ,
\end{equation}
which can be computed using the Baker--Campbell--Hausdorff (BCH) expansion and can thus be viewed as a perturbation of the exact Fokker--Planck operator $\mathcal{L}^{\dag}$:
\begin{equation}\label{eq:hat_L}
  \hat{\mathcal{L}}^{\dag} = \mathcal{L}^{\dag} + h\mathcal{L}^{\dag}_{1} + h^{2}\mathcal{L}^{\dag}_{2} + O(h^{3})
\end{equation}
for some perturbation operators $\mathcal{L}^{\dag}_{i}$.

We also define the invariant distribution $\hat{\rho}$ associated with the numerical method as an approximation of the target invariant distribution $\rho_{\beta}$:
\begin{equation}\label{eq:hat_rho}
  \hat{\rho} = \rho_{\beta}\left[ 1+hf_{1}+h^{2}f_{2}+h^{3}f_{3}+O(h^{4}) \right]
\end{equation}
for some correction functions $f_{i}$ satisfying $\langle f_{i} \rangle=0$, where $\langle \cdot \rangle$ denotes the average with respect to the target invariant distribution.

Thus, substituting $\hat{\mathcal{L}}^{\dag}$ and $\hat{\rho}$ into the stationary Fokker--Planck equation
\begin{equation}
  \hat{\mathcal{L}}^{\dag}\hat{\rho} = 0
\end{equation}
gives
\begin{equation}
  \left( \mathcal{L}^{\dag} + h\mathcal{L}^{\dag}_{1} + h^{2}\mathcal{L}^{\dag}_{2} + O(h^{3}) \right)\left(\rho_{\beta}\left[1+hf_{1}+h^{2}f_{2}+h^{3}f_{3}+O(h^{4}) \right]\right)=0 \, .
\end{equation}
Since the exact Fokker--Planck operator preserves the target invariant distribution, i.e., $\mathcal{L}^{\dag}\rho_{\beta}=0$, we obtain
\begin{equation}\label{eq:error_analysis_PDE}
  \mathcal{L}^{\dag}(\rho_{\beta}f_{1}) = - \mathcal{L}^{\dag}_{1}\rho_{\beta}
\end{equation}
by equating first order terms in $h$.

We are able to find the perturbation operator $\mathcal{L}^{\dag}_{1}$ by using the BCH expansion for any particular integration scheme.  Then we can calculate its action on $\rho_{\beta}$.  However, it is generally very hard to solve the above PDE~\eqref{eq:error_analysis_PDE} in order to obtain the leading correction function $f_{1}$ (see examples in Langevin dynamics~\cite{Leimkuhler2013}). Note that for symmetric splitting methods, including the PAdL scheme~\eqref{eq:PAdL-S}, $f_{1}$ is always equal to zero, thereby ensuring second order convergence to the invariant measure~\cite{Leimkuhler2015a}.

Consider now the PNHL-N method~\eqref{eq:PNHL-N}, which can be written as
\begin{equation}
  \exp\left(h\hat{\mathcal{L}}^{\dag}_\mathrm{PNHL-N}\right) = \exp\left(\frac{h}{2}\mathcal{L}^{\dag}_\mathrm{X}\right) \exp\left(h\mathcal{L}^{\dag}_\mathrm{Y}\right) \exp\left(\frac{h}{2}\mathcal{L}^{\dag}_\mathrm{X}\right) \, ,
\end{equation}
where
\begin{equation}
  \exp\left(\frac{h}{2}\mathcal{L}^{\dag}_\mathrm{X}\right) = \exp\left(\frac{h}{2}\mathcal{L}^{\dag}_\mathrm{B}\right) \exp\left(\frac{h}{2}\mathcal{L}^{\dag}_\mathrm{A}\right) \, ,
\end{equation}
and
\begin{equation}
  \exp\left(h\mathcal{L}^{\dag}_\text{Y}\right) = \exp\left(\frac{h}{2}\mathcal{L}^{\dag}_\text{C}\right) \exp\left(\frac{h}{2}\mathcal{L}^{\dag}_\text{D}\right) \exp\left(h\mathcal{L}^{\dag}_\text{O}\right)
  \exp\left(\frac{h}{2}\mathcal{L}^{\dag}_\text{D}\right) \exp\left(\frac{h}{2}\mathcal{L}^{\dag}_\text{C}\right) \, .
\end{equation}

By using the BCH expansion, we obtain
\begin{equation}
  \begin{aligned}
    \mathcal{L}^{\dag}_\mathrm{X} &= \mathcal{L}^{\dag}_\mathrm{A} + \mathcal{L}^{\dag}_\mathrm{B} - \frac{h}{4} \left[\mathcal{L}^{\dag}_\mathrm{A},\mathcal{L}^{\dag}_\mathrm{B}\right] + O(h^{2}) \, , \\
    \mathcal{L}^{\dag}_\mathrm{Y} &= \mathcal{L}^{\dag}_\mathrm{C} + \mathcal{L}^{\dag}_\mathrm{D} + \mathcal{L}^{\dag}_\mathrm{O} + O(h^{2}) \, ,
  \end{aligned}
\end{equation}
where the notation $[\mathcal{A},\mathcal{B}]=\mathcal{A}\mathcal{B}-\mathcal{B}\mathcal{A}$ denotes the commutator of operators $\mathcal{A}$ and $\mathcal{B}$, and subsequently
\begin{equation}
  \hat{\mathcal{L}}^{\dag}_\mathrm{PNHL-N} = \mathcal{L}^{\dag}_\mathrm{A} + \mathcal{L}^{\dag}_\mathrm{B} + \mathcal{L}^{\dag}_\mathrm{C} + \mathcal{L}^{\dag}_\mathrm{D} + \mathcal{L}^{\dag}_\mathrm{O} - \frac{h}{4} \left[\mathcal{L}^{\dag}_\mathrm{A},\mathcal{L}^{\dag}_\mathrm{B}\right] + O(h^{2}) \, .
\end{equation}
Thus the leading perturbation operator of the PNHL-N scheme is
\begin{equation}\label{eq:L1_PNHL-N}
  \mathcal{L}^{\dag}_\mathrm{1,PNHL-N} = - \frac{1}{4} \left[\mathcal{L}^{\dag}_\mathrm{A},\mathcal{L}^{\dag}_\mathrm{B}\right] \, ,
\end{equation}
whose action on the invariant distribution of the PNHL system~\eqref{eq:Gibbs PNHL} reads (assuming $\M=\I$ for simplicity)
\begin{equation}
  \mathcal{L}^{\dag}_\mathrm{1,PNHL-N}\hat{\rho}_{\beta}(\q,\p,\xi) = - \frac{\beta}{4} \left( \mathbf{p}^{T} \Delta U(\mathbf{q})\mathbf{p} - [\nabla U(\mathbf{q})]^{T}\nabla U(\mathbf{q}) \right)\hat{\rho}_{\beta} \, .
\end{equation}

Recall the Fokker--Planck operator of the PNHL system:
\begin{equation}
  \mathcal{L}^{\dag}_\mathrm{PNHL} = - \mathbf{p} \cdot \nabla_{\mathbf{q}} + \nabla U(\mathbf{q}) \cdot \nabla_{\mathbf{p}} + \xi \nabla_{\mathbf{p}} \cdot (\boldsymbol{\Gamma}(\mathbf{q})\mathbf{p}\cdot ) - G(\mathbf{q},\mathbf{p}) \frac{\partial}{\partial\xi} + \tilde{\gamma}\frac{\partial}{\partial\xi}(\xi\cdot ) + \frac{{\tilde{\sigma}}^{2}}{2}\frac{\partial^{2}}{\partial\xi^{2}} \, .
\end{equation}
Although in this case the right-hand side of the PDE~\eqref{eq:error_analysis_PDE} is relatively simple, it is still very challenging to solve the PDE in order to obtain the corresponding leading correction function $f_\mathrm{1,PNHL-N}$. However, the additional variable $\xi$ is normally distributed with mean zero and variance $\beta^{-1}\mu^{-1}$. Thus, the variance of $\xi$ will be small if the thermal mass $\mu$ is large. Therefore, following~\cite{Leimkuhler2015a}, we consider projecting the Fokker--Planck equation and its solution by integrating with respect to the Gaussian distribution of $\xi$ in the ergodic limit. That is, we apply the projection operator~\cite{Givon2004}
\begin{equation}
  \mathcal{P} \nu(\q,\p,\xi) := \frac{\int_{\Omega_\xi} \hat{\rho}_{\beta}(\q,\p,\xi) \nu(\q,\p,\xi) \, {\rm d} \xi}{\int_{\Omega_\xi} \hat{\rho}_{\beta}(\q,\p,\xi) \, {\rm d} \xi} \, ,
\end{equation}
where $\nu$ is an arbitrary function, to the PDE~\eqref{eq:error_analysis_PDE}.  Effectively, this leads to the reduced equation
\begin{equation}\label{eq:error_analysis_PDE_Ad-Lreduced}
  \check{\mathcal{L}}^{\dag}({\rho}_{\beta}\hat{f}_{1}) = - \rho_{\beta} \mathcal{P} \frac{\mathcal{L}^{\dag}_{1}\hat{\rho}_{\beta}}{\hat{\rho}_{\beta}} \, ,
\end{equation}
where the operator $\check{\mathcal{L}}^{\dag}$, acting on functions of $\q$ and $\p$ only, is just the operator $\mathcal{L}^{\dag}$ reduced by the action of the projection. In fact, now $\mathcal{L}^{\dag}_\mathrm{PNHL}$ is simply reduced to
\begin{equation}
  \check{\mathcal{L}}^{\dag}_\mathrm{PNHL} = - \mathbf{p} \cdot \nabla_{\mathbf{q}} + \nabla U(\mathbf{q}) \cdot \nabla_{\mathbf{p}} \, ,
\end{equation}
while the right-hand side is unchanged due to the fact that $\xi$ is not present there. Finally, we can solve the reduced PDE to obtain the leading correction function:
\begin{equation}
  \hat{f}_\mathrm{1,PNHL-N} =  \frac{\beta}{4}\mathbf{p}^{T}\nabla U(\mathbf{q}) \, ,
\end{equation}
which leads to the following proposition:
\\
\par\noindent {\bf Proposition 1.} \emph{For sufficiently large thermal mass, the PNHL-N method~\eqref{eq:PNHL-N} exhibits second order convergence for thermodynamic averages of certain observables, i.e., in the form of $\phi(\mathbf{q},\mathbf{p}) = \mathbf{p}^{2k}\vartheta(\mathbf{q})$, where $k$ is an integer and $\vartheta(\mathbf{q})$ can be constant.}
\\
\\
The class of observables includes kinetic temperature and observables that only depend on the configurations. In other words, for those observables, assuming the asymptotic expansion holds, the computed average (in the large thermal mass limit) reads
\begin{equation}
  \langle\phi\rangle_{h} = \langle\phi\rangle + h\langle\phi \hat{f}_{1}\rangle + h^{2}\langle\phi \hat{f}_{2}\rangle + \dots = \langle\phi\rangle + O(h^{2}) \, ,
\end{equation}
which is of order two. This is fully consistent with what we have observed numerically for a number of observables in~\cite{Leimkuhler2015} and which we further verify in the following section.

\section{Numerical experiments}
\label{sec:Numerical_Experiments}

In this section, we conduct various numerical experiments, in both equilibrium and nonequilibrium regimes, to compare the newly proposed PAdL method with a number of alternative momentum-conserving schemes described in~\cite{Leimkuhler2015}.

\subsection{Simulation details}

\begin{figure}[tb]
\centering
\includegraphics[scale=0.5]{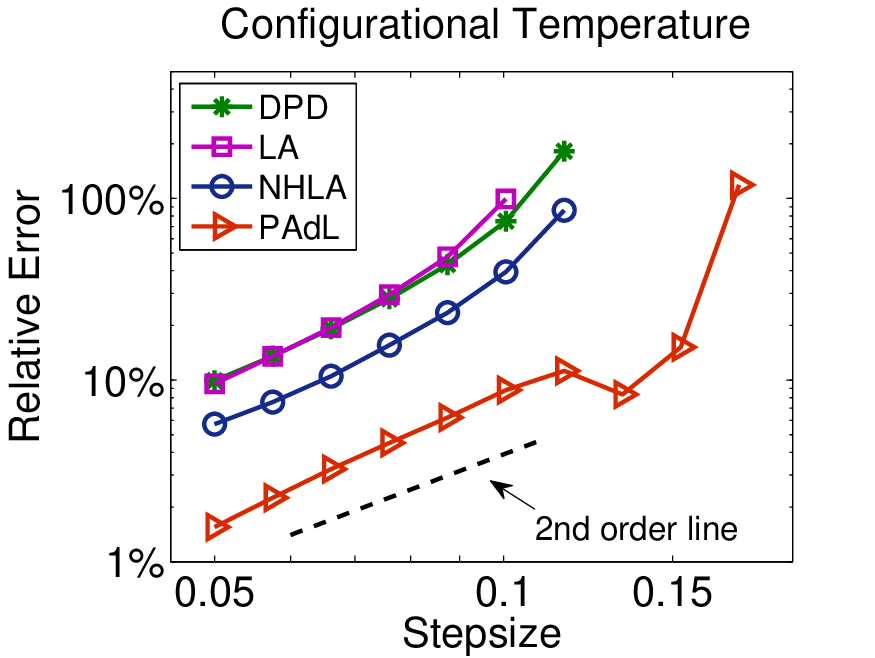}
\includegraphics[scale=0.5]{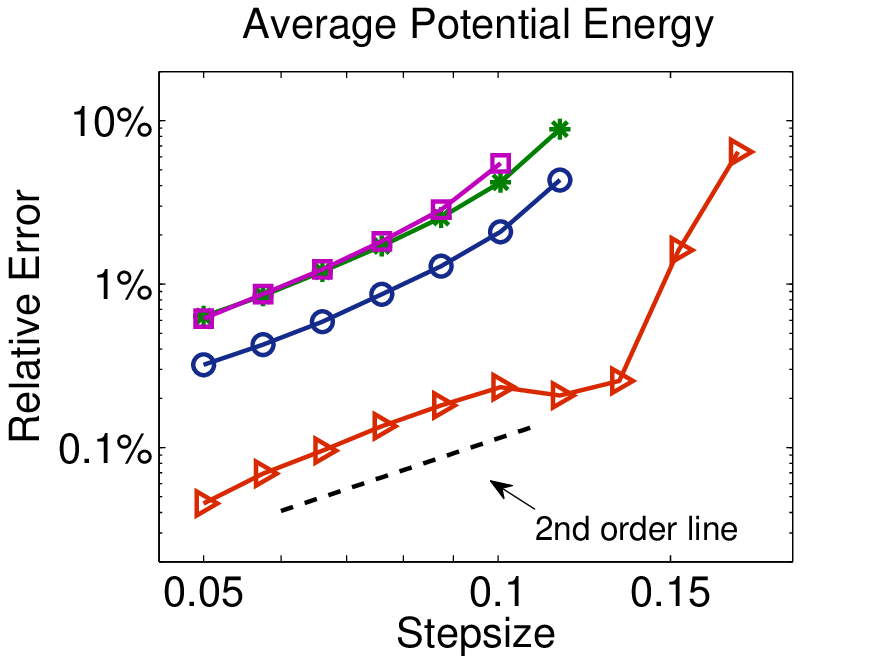}
\caption{\small Log-log plot of the relative error in computed configurational temperature (left) and average potential energy (right) against stepsize by using various numerical methods with (effective) friction coefficient $\gamma=4.5$. The system was simulated for 1000 reduced time units but only the last 80\% of the data were collected to calculate the static quantity to make sure the system was well equilibrated. Ten different runs were averaged to further reduce the sampling errors. The stepsizes tested began at $h=0.05$ and were increased incrementally by 15\% until all methods either started to show significant relative error (100\% in configurational temperature or 10\% in average potential energy) or became unstable.}
\label{fig:PAdL_CT_U_Comp_gamma4d5}
\end{figure}

\begin{table}[tb]
\begin{center}
%\fontsize{9}{10.8}\selectfont{
\resizebox{1.0\textwidth}{!}{\begin{minipage}{\textwidth}
    \begin{tabular}{ |c|c|c|c|c| }
    \hline
    \textbf{Method} & \textbf{Critical stepsize} & \textbf{Maximal stepsize} &% \textbf{Force Calculation} &
     \textbf{CPU time} & \textbf{Scaled efficiency} \\ \hline
    DPD-VV      & 0.05 & 0.10 & %1 &
    19.878 & 100.0\% \\ \hline
    DPD-S1      & 0.05 & 0.11 & %1 &
     20.018 & 99.3\% \\ \hline
    DPD-Trotter & 0.05 & 0.11 & %1 &
    20.788 & 95.6\% \\ \hline
    Peters      & 0.05 & 0.11 & %1 &
    20.893 & 95.1\% \\ \hline
    LA          & 0.05 & 0.10 & %1 &
    17.808 & 111.6\% \\ \hline
    NHLA        & 0.07 & 0.13 & %1 &
    18.513 & 150.3\% \\ \hline
    PAdL        & 0.13 & 0.17 & %1 &
    22.219 & 232.6\% \\
    \hline
    \end{tabular}
\caption[Table caption text]{\small Comparisons of the computational efficiency of the various numerical methods in the moderate (effective) friction regime of $\gamma=4.5$. ``Critical stepsize'' is the stepsize beyond which the numerical method starts to show pronounced artifacts (10\% relative error in computed configurational temperature), while ``maximal stepsize'' is the stepsize stability threshold above which the method is unstable. The ``numerical efficiency'' of each method was scaled to that of the benchmark DPD-VV method. The efficiency figures above quantify the computational work in terms of the number of force computations and correctly take into account the fact that all the methods require one force evaluation per timestep with the sole exception of PNHL-N (included only in the low friction case, see Table~\ref{table:efficiency_eq_gamma_0d5}) which requires two force evaluations.  }
\label{table:efficiency_eq_gamma_4d5}
\end{minipage} } %}
\end{center}
\end{table}

We adopted the simulation details used in~\cite{Leimkuhler2015}, a standard parameter set commonly used in algorithms tests~\cite{Groot1997,Vattulainen2002,Nikunen2003,Shardlow2003,Peters2004,Stoyanov2005,Serrano2006}. Specifically, a system of $N=500$ identical particles with unity mass was simulated in a cubic box (length $L=5$) with periodic boundary conditions, unless otherwise stated. Particle density $\rho_{\rm d}=4$ was used with cutoff radius $r_{\mathrm{c}}=1$ and $k_{\mathrm{B}}T=1$. The initial positions of the particles were independent and identically distributed (i.i.d.) with a uniform distribution over the box, while the initial momenta were i.i.d.\ normal random variables with mean zero and variance $k_{\mathrm{B}}T$. The potential repulsion parameters $a_{ij}$ were set to be 25, while a wide range of (effective) friction coefficients (0.5, 4.5, and 40.5) were used. Verlet neighbor lists~\cite{Verlet1967} were used in each method.

In our simulations, the thermal mass in PAdL was chosen to be the same as that of PNHL (where $\tilde{\gamma}=4.5$ was used), i.e., $\mu=10$. When comparing different formulations, we have to make sure that similar translational diffusion properties of the fluid were obtained. For the PAdL thermostat, we can always tune the value of $\sigma$ so that the same (effective) friction coefficient as in DPD was obtained, i.e., $\hat{\gamma}=\gamma$. For both the Lowe--Andersen (LA) thermostat~\cite{Lowe1999} and the Nos\'{e}--Hoover--Lowe--Andersen (NHLA) thermostat~\cite{Stoyanov2005}, the stochastic randomization frequency $\Gamma$ was set to be 0.44 as in~\cite{Nikunen2003,Jakobsen2005}, which corresponds to the case of $\gamma=4.5$ in DPD. For cases of $\gamma=0.5$ and $\gamma=40.5$ in DPD, we used $\Gamma=0.1$ and $\Gamma=4$, respectively.

We have observed in~\cite{Leimkuhler2015} that standard DPD methods (including the DPD velocity Verlet method (DPD-VV)~\cite{Besold2000}, Shardlow's S1 splitting method (DPD-S1)~\cite{Shardlow2003}, and the DPD Trotter scheme (DPD-Trotter)~\cite{Serrano2006}) and the Peters thermostat~\cite{Peters2004} give almost indistinguishable performance in all the quantities that we have tested. Therefore, the DPD-S1 method was used to represent the standard DPD formulation, unless otherwise stated. The PNH thermostat~\cite{Allen2007} was not included for further comparison because of its stability issue at relatively small stepsize, which is well documented in~\cite{Leimkuhler2015}. Since the dynamics of the PNHL thermostat is consistent with that of DPD in the low friction regime, both PNHL methods were compared to alternatives only in the case of $\gamma=0.5$.

As in~\cite{Leimkuhler2015}, we measured the ``numerical efficiency'', defined as the ratio of the ``critical stepsize'' and CPU time per step, of each method and then scaled it to the benchmark DPD-VV method, which is widely used in popular software packages. The CPU time (in milliseconds) for the main integration step (without calculating any physical quantity) is the time taken (on an HP Z600 Workstation with 15.7 GB RAM) with the use of Verlet neighbor lists for the integration of a single time step of $h=0.05$ (averaged over 10000 consecutive time steps). The critical stepsize was chosen as the stepsize corresponding to 10\% relative error in computed configurational temperature~\cite{Rugh1997,Braga2005,Allen2006,Travis2008}, an observable function of positions whose average in the canonical ensemble is precisely the target temperature:
\begin{equation}\label{eq:Config_Temp}
  k_{\mathrm{B}}T = \frac{\sum_{i}\langle {\| \nabla_{i}U \|}^{2} \rangle}{\sum_{i}\langle \nabla^{2}_{i}U \rangle} \, ,
\end{equation}
where $\nabla_{i}U$ and $\nabla^{2}_{i}U$ denote the gradient and Laplacian of the potential energy $U$ with respect to the position of particle $i$, respectively (see more discussions in~\cite{Leimkuhler2015}). We mention that this criterion was not precisely adopted in a recent article by Farago and Gr{\o}nbech-Jensen~\cite{Farago2016}, where they determined the critical stepsize based on the accuracy control of average potential energy but compared their results to those obtained from configurational temperature in~\cite{Leimkuhler2015}. We would also like to point out that the Nos\'{e}--Hoover device employed in PAdL and PNHL is straightforward to implement.  In MD simulations, the choice of the thermal mass $\mu$ can be a technical impediment to using Nos\'{e}--Hoover controls, but we have observed in~\cite{Leimkuhler2009} and also in the experiments of this article that the addition of stochastic noise changes the nature of the coupling parameter and the results obtained are relatively insensitive to its selection.

\subsection{Equilibrium}

\begin{figure}[tb]
\centering
\includegraphics[scale=0.5]{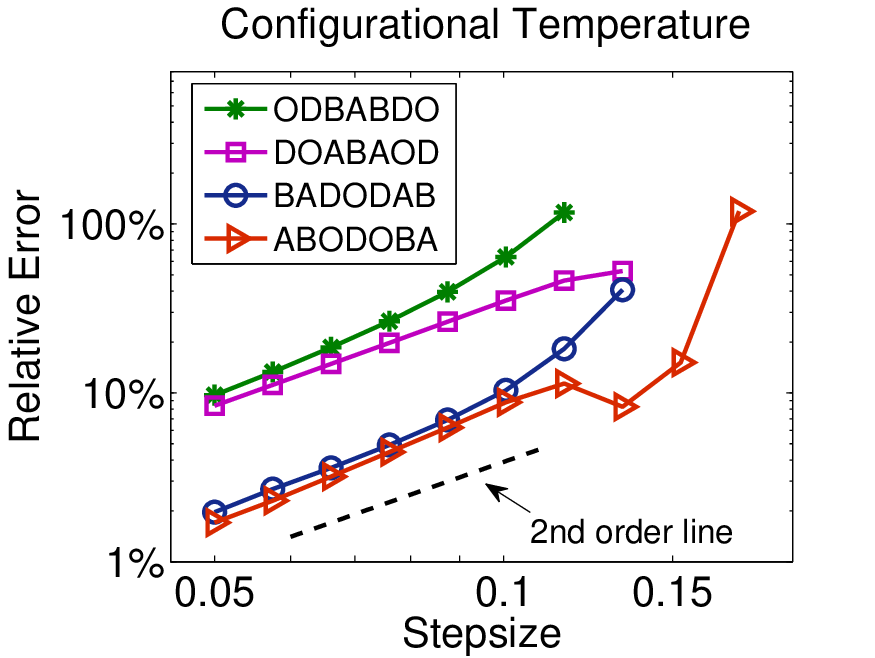}
\caption{\small Comparisons of the relative error in computed configurational temperature against stepsize by using various splitting methods of the PAdL system with effective friction coefficient $\hat{\gamma}=4.5$. The format of the plots is the same as in Fig.~\ref{fig:PAdL_CT_U_Comp_gamma4d5}. }
\label{fig:PAdL_CT_Comp_gamma4d5_Methods}
\end{figure}

Configurational quantities, such as configurational temperature and average potential energy, were compared in Figure~\ref{fig:PAdL_CT_U_Comp_gamma4d5} with (effective) friction coefficient $\gamma=4.5$. According to the dashed order line, we can see that all the methods tested exhibit second order convergence to the invariant measure for both quantities. More specifically, DPD and the LA thermostat show rather similar behavior, while the NHLA thermostat is slightly better than those two. Quite remarkably, the newly proposed PAdL method~\eqref{eq:PAdL-S} achieves one order of magnitude improvement over DPD in terms of numerical accuracy for a fixed stepsize. For certain accuracy (i.e., a fixed relative error), the PAdL method can use doubled stepsize, thus substantially improving the ``numerical efficiency'' defined in~\cite{Leimkuhler2015} (100\% efficiency improvement = double the performance). Our observations were confirmed in Table~\ref{table:efficiency_eq_gamma_4d5}, which shows that
the PAdL method indeed has a more than 130\% improvement in numerical efficiency over the DPD method. The results on the configurational temperature and average potential energy are rather similar, therefore in what follows we present only configurational temperature results.

\begin{figure}[tb]
\centering
\includegraphics[scale=0.5]{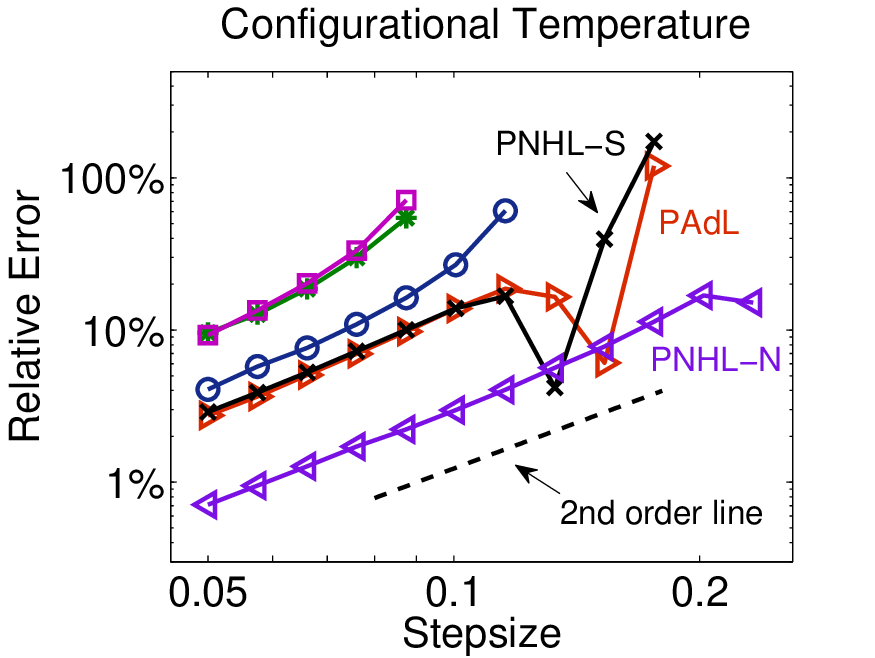}
\includegraphics[scale=0.5]{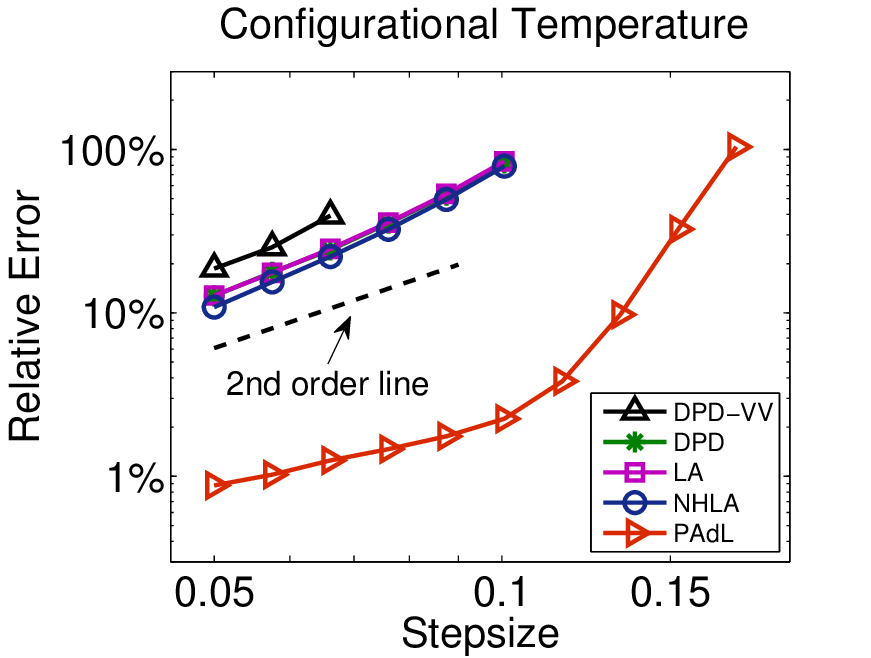}
\caption{\small Comparisons of the relative error in computed configurational temperature against stepsize by using various numerical methods with (effective) friction coefficient $\gamma=0.5$ (left) and $\gamma=40.5$ (right). Note that two PNHL methods, which correspond to the low friction regime of DPD, were included in the former case only. The format of the plots is the same as in Fig.~\ref{fig:PAdL_CT_U_Comp_gamma4d5}. }
\label{fig:PAdL_CT_Comp_gamma0d5_40d5}
\end{figure}

\begin{table}[tb]
\begin{center}
%\fontsize{11}{13.2}\selectfont{
\resizebox{1.0\textwidth}{!}{\begin{minipage}{\textwidth}
    \begin{tabular}{ |c|c|c|c|c| }
    \hline
    \textbf{Method} & \textbf{Critical stepsize} & \textbf{Maximal stepsize} & %\textbf{Force calculation} &
     \textbf{CPU time} & \textbf{Scaled efficiency} \\ \hline
    DPD-VV      & 0.05 & 0.09 & %1 &
    19.878 & 100.0\% \\ \hline
    DPD-S1      & 0.05 & 0.09 & % 1 &
    20.018 & 99.3\% \\ \hline
    DPD-Trotter & 0.05 & 0.09 & %1 &
    20.788 & 95.6\% \\ \hline
    Peters      & 0.05 & 0.09 & %1 &
    20.893 & 95.1\% \\ \hline
    LA          & 0.05 & 0.09 & %1 &
    17.808 & 111.6\% \\ \hline
    NHLA        & 0.07 & 0.11 & %1 &
    18.513 & 150.3\% \\ \hline
    PNH         & 0.05 & 0.08 & %1 &
    16.450 & 120.8\% \\ \hline
    PNHL-S      & 0.08 & 0.17 & %1 &
    21.199 & 150.0\% \\ \hline
    PNHL-N      & 0.17 & 0.23 & % 2 &
    37.206 & 181.7\% \\ \hline
    PAdL        & 0.08 & 0.17 & % 1 &
    22.219 & 143.1\% \\
    \hline
    \end{tabular}
\caption[Table caption text]{\small Comparisons of the computational efficiency of the various numerical methods in the low (effective) friction regime of $\gamma=0.5$. The format of the table is the same as in Table~\ref{table:efficiency_eq_gamma_4d5}. }
\label{table:efficiency_eq_gamma_0d5}
\end{minipage} } %}
\end{center}
\end{table}

\begin{table}[tb]
\begin{center}
%\fontsize{9}{10.8}\selectfont{
\resizebox{1.0\textwidth}{!}{\begin{minipage}{\textwidth}
    \begin{tabular}{ |c|c|c|c|c| }
    \hline
    \textbf{Method} & \textbf{Critical stepsize} & \textbf{Maximal stepsize} & %\textbf{Force calculation} &
     \textbf{CPU time} & \textbf{Scaled efficiency} \\ \hline
    DPD-VV      & 0.04 & 0.07 & %1 &
    19.878 & 100.0\% \\ \hline
    DPD-S1      & 0.05 & 0.11 & %1 &
    20.018 & 124.1\% \\ \hline
    DPD-Trotter & 0.03 & 0.08 & %1 &
    20.788 & 119.5\% \\ \hline
    Peters      & 0.05 & 0.11 & %1 &
    20.893 & 118.9\% \\ \hline
    LA          & 0.05 & 0.11 & %1 &
    17.808 & 139.5\% \\ \hline
    NHLA        & 0.05 & 0.11 & %1 &
    18.513 & 134.2\% \\ \hline
    PAdL        & 0.13 & 0.17 & %1 &
    22.219 & 290.8\% \\
    \hline
    \end{tabular}
\caption[Table caption text]{\small Comparisons of the computational efficiency of the various numerical methods in the high (effective) friction regime of $\gamma=40.5$. The format of the table is the same as in Table~\ref{table:efficiency_eq_gamma_4d5}.}
\label{table:efficiency_eq_gamma_40d5}
\end{minipage} } %}
\end{center}
\end{table}

We also explore in Figure~\ref{fig:PAdL_CT_Comp_gamma4d5_Methods} the performance of various splitting methods of the PAdL system with a fixed value of $\hat{\gamma}=4.5$. All the methods compared clearly display second order convergence to the invariant measure, with ABODOBA and BADODAB achieving one order of magnitude improvement in accuracy compared to the other methods. However, it should be noted that in ABODOBA, we needed to update the Verlet neighbor lists only once at each step since the update of A at the end was directly followed by another update of A at the beginning of next step, whereas one additional update of Verlet neighbor lists was required in BADODAB. In our simulation code, the cost of such an update was essentially the same as one force calculation, thus the ABODOBA method~\eqref{eq:PAdL-S} was used throughout the current article for the PAdL system. This again illustrates the importance of optimal design of numerical methods.

Figure~\ref{fig:PAdL_CT_Comp_gamma0d5_40d5} (left) compares the configurational temperature control of various methods in both low and high friction regimes. In the low friction regime, where both PNHL methods were included, again all the methods exhibit second order convergence to the invariant measure. The NHLA, PNHL-S, and PAdL methods are rather similar to each other, all of which are superior to both DPD and LA methods. The PNHL-N method achieves more than one order of magnitude improvement in numerical accuracy over the DPD method. Although the PNHL-N method requires two force calculations at each step, it still achieves a more than 80\% improvement as shown in Table~\ref{table:efficiency_eq_gamma_0d5}. The table also reveals that the PAdL method has an almost 50\% improvement in performance compared to the DPD method.

In the high friction regime, the behavior of those methods are rather different from that of other regimes. As shown in Figure~\ref{fig:PAdL_CT_Comp_gamma0d5_40d5} (right), surprisingly the most popular DPD-VV method is slightly worse than other standard DPD methods. Both LA and NHLA are indistinguishable from the DPD method. Superconvergence property (i.e., fourth order convergence to the invariant measure) was not observed for the PAdL method in this regime. Nevertheless, the PAdL method still obtains a dramatic improvement over all other schemes. Table~\ref{table:efficiency_eq_gamma_40d5} shows that the PAdL method has a more than 190\% improvement in overall numerical efficiency over the benchmark DPD method.

\begin{figure}[htb]
\centering
\includegraphics[scale=0.5]{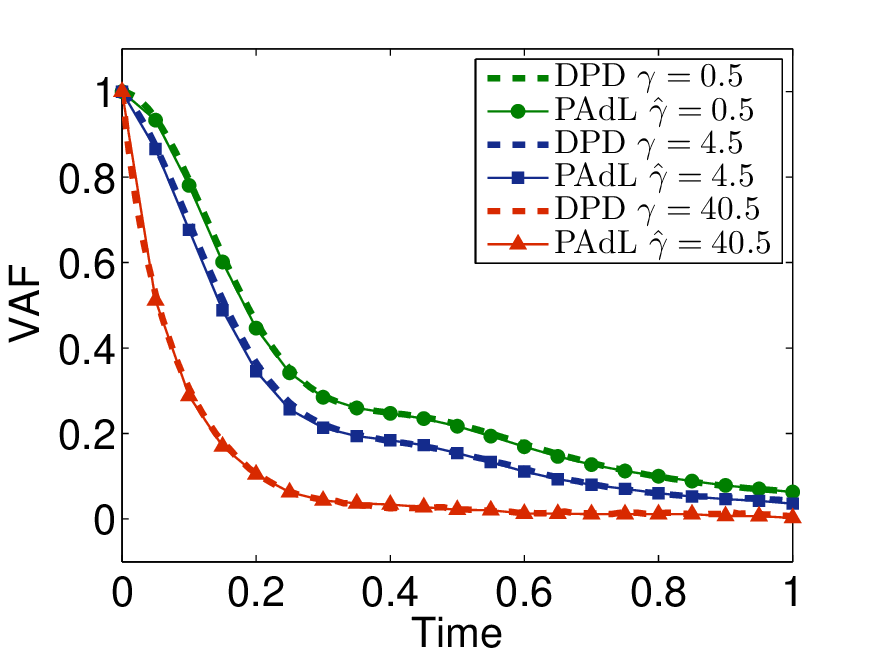}
\includegraphics[scale=0.5]{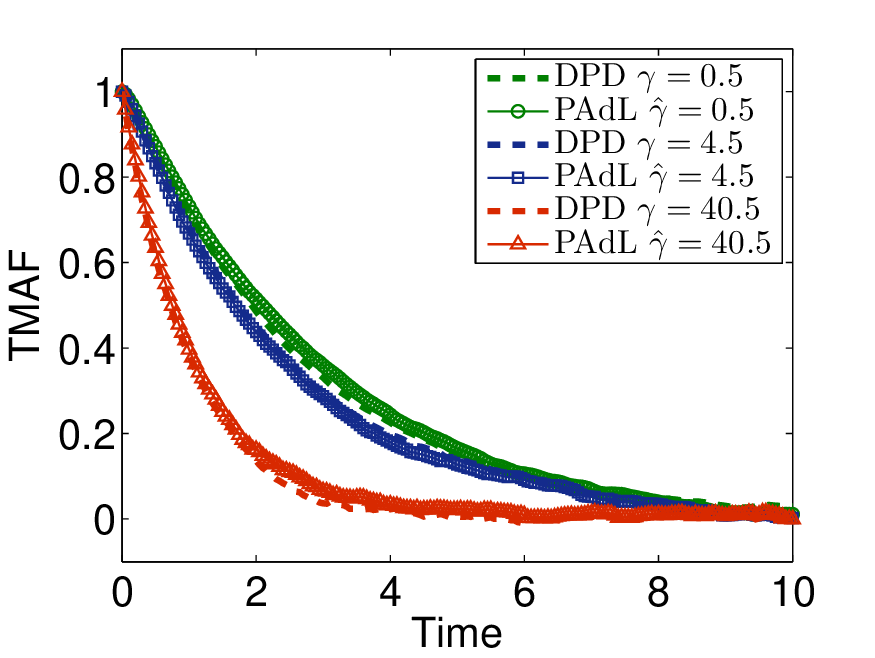}
\caption{\small (Color.) Comparisons of velocity autocorrelation function (VAF) (left) and transverse momentum autocorrelation function (TMAF) (right) between the standard DPD method and the newly established PAdL scheme with three different values of the (effective) friction coefficient. The DPD-S1 method was used for DPD with a small stepsize of $h=0.01$, while $h=0.05$ was used for PAdL. 100 and 100,000 different runs were averaged in the cases of VAF and TMAF (the wavenumber was chosen as $2\pi/L$), respectively, to reduce the sampling errors after the system was well equilibrated.}
\label{fig:PAdL_VAF_Comp}
\end{figure}

The control of the dynamical properties of the PAdL method was also tested and plotted in Figure~\ref{fig:PAdL_VAF_Comp}. In particular, we compared two important quantities: the velocity autocorrelation function (VAF) and the transverse momentum autocorrelation function (TMAF)~\cite{Tang1995,Hansen2006}, which characterize the translational and rotational diffusions of the system, respectively. The integral of the VAF is related to the diffusion constant, while the logarithm of the TMAF is proportional to the shear viscosity in the hydrodynamic limit (see more discussions in~\cite{Leimkuhler2015}). Unlike the PNHL formulation, which corresponds to the low friction regime, the PAdL system is able to capture the dynamics of DPD in a wide range of friction coefficients---the relaxations of the VAF and the TMAF of both formulations are indeed indistinguishable (only visible with the help of the on-center symbols).

\subsection{Nonequilibrium}

It is well known that nonequilibrium methods, where a steady state is maintained under external perturbations (either stationary fluxes or spatial gradients of some quantities), are more efficient means than equilibrium autocorrelation functions for extracting transport coefficients (e.g., rheological properties such as shear stress and shear viscosity) from fluid dynamics simulations (see more discussions in~\cite{Allen1989}). Thus, there has been a rapidly growing interest in NEMD~\cite{Evans2008}. For instance, planar Couette flow, where a simple and steady shear flow is generated, is commonly employed as a numerical ``viscometer'' in particle-based methods to obtain transport coefficients~\cite{Allen1989,Evans2008}. Furthermore, due to the fact that it is relevant to many real-life phenomena as well as for its simplicity, planar Couette flow has been widely adopted in laboratory experiments.

In order to measure the shear viscosity in lab experiments, a linear profile is often imposed at a fixed shear rate and then the resultant shear stress can be measured. However, in computer simulations, simple periodic boundary conditions are unable to maintain a steady linear velocity profile, resulting in problems at the boundaries of the simulation domain. There have been early attempts to generate momentum or energy flows in MD simulations where particles are made to interact with momentum or energy reservoirs (e.g., a velocity profile can be obtained by fixing the average velocity in the extremal slabs of a fluid)~\cite{Ashurst1975,Ciccotti1980,Trozzi1984,Joubaud2012}. However, these methods are not compatible with periodic boundary conditions, and thus lead to surface effects. Alternatively and more appealingly, one can apply Lees--Edwards boundary conditions (LEBC)~\cite{Lees1972} to retain periodicity but alter the position and velocity of the periodic images. In this case, a simple shear flow (with a constant shear rate) is generated, which allows the investigation of the dependence of the viscosity on the shear rate~\cite{Fedosov2011}. LEBC has been extensively studied in DPD and related systems to study rheological behavior in colloidal suspensions~\cite{Boek1996,Boek1997}, polymeric systems~\cite{Soddemann2003,Pastorino2007,Fedosov2010}, as well as multiphase systems~\cite{Pan2014} (see more discussions on boundary conditions in DPD in~\cite{Revenga1999,Pivkin2005,Pivkin2010}).

Before we analyze the numerical results obtained by simulating various methods under LEBC (i.e., the system is perturbed by a simple shear flow), we first briefly review LEBC and then discuss two important issues in NEMD: (1) the practical implementation of LEBC in DPD and related momentum-conserving systems, where forces are dependent on relative velocities; (2) the practical measurement of system temperature in NEMD.

\subsubsection{Lees--Edwards boundary conditions (LEBC)}

In order to generate a simple shear flow in NEMD, the periodic boundary conditions (PBC) have to be modified. A common way to achieve that is to apply the well-known LEBC~\cite{Lees1972}. In LEBC, the primary cubic box (with lengths $L_{x}$, $L_{y}$, and $L_{z}$) remains centered at the origin, however, a uniform shear velocity profile is expected~\cite{Evans2008}
\begin{equation}\label{eq:LEBC_Streaming_V}
  \mathbf{u} = \dot{\gamma}y\mathbf{e}^{x} \, ,
\end{equation}
where $\mathbf{e}^{x}$ is the unit vector in the $x$-direction and $\dot{\gamma}$ is the shear rate defined as
\begin{equation}
  \dot{\gamma} = \frac{\dd v^{x}}{\dd y} \, .
\end{equation}
LEBC is also called the ``sliding brick'' boundary conditions. It is important to note that LEBC is applied only in the $x$-direction, while the other directions ($y$ and $z$) remain with PBC.

Special attention has to be paid in LEBC when a particle is crossing the boundary in the $y$-direction. In this case, one of the images of the crossing particle will enter through the opposite face, but with both position and velocity modified in a proper way because of the streaming velocity~\eqref{eq:LEBC_Streaming_V}.

The periodic boundary crossing is now handled as follows~\cite{Allen1989}:
\begin{align}
  N_{\mathrm{L}} &= \mathrm{round}(q^{y}_{i}/L_{y}) \\
  q^{x}_{i} &\leftarrow q^{x}_{i} - N_{\mathrm{L}} \Delta q^{x} \\
  q^{x}_{i} &\leftarrow q^{x}_{i} - L_{x} \cdot \mathrm{round} (q^{x}_{i}/L_{x}) \\
  q^{y}_{i} &\leftarrow q^{y}_{i} - L_{y} N_{\mathrm{L}} \\
  q^{z}_{i} &\leftarrow q^{z}_{i} - L_{z} \cdot \mathrm{round} (q^{z}_{i}/L_{z}) \\
  v^{x}_{i} &\leftarrow v^{x}_{i} - N_{\mathrm{L}} \dot{\gamma}L_{y}
\end{align}
where $N_{\mathrm{L}}$ is the ``rounded'' number of layers (boxes) between the current position of particle $i$ in the $y$-direction and the origin, and $\Delta q^{x}$ is the displacement of the upper layer during the elapsed time $t$ from an
appropriate origin, i.e.,
\begin{equation}
  \Delta q^{x} = \dot{\gamma}L_{y}t \, .
\end{equation}

The minimum image convention should now proceed as follows~\cite{Allen1989}:
\begin{align}
  N_{\mathrm{L}} &= \mathrm{round}(q^{y}_{ij}/L_{y}) \label{eq:LEBC_MIC-1} \\
  q^{x}_{ij} &\leftarrow q^{x}_{ij} - N_{\mathrm{L}} \Delta q^{x} \label{eq:LEBC_MIC-2} \\
  q^{x}_{ij} &\leftarrow q^{x}_{ij} - L_{x} \cdot \mathrm{round} (q^{x}_{ij}/L_{x}) \label{eq:LEBC_MIC-3} \\
  q^{y}_{ij} &\leftarrow q^{y}_{ij} - L_{y} N_{\mathrm{L}} \label{eq:LEBC_MIC-4} \\
  q^{z}_{ij} &\leftarrow q^{z}_{ij} - L_{z} \cdot \mathrm{round} (q^{z}_{ij}/L_{z}) \label{eq:LEBC_MIC-5}
\end{align}

Note that when the shear rate is zero, i.e., $\dot{\gamma} = 0$, LEBC reduces to PBC.

\subsubsection{LEBC in pairwise thermostats}

\begin{figure}[tb]
\centering
\includegraphics[scale=0.5]{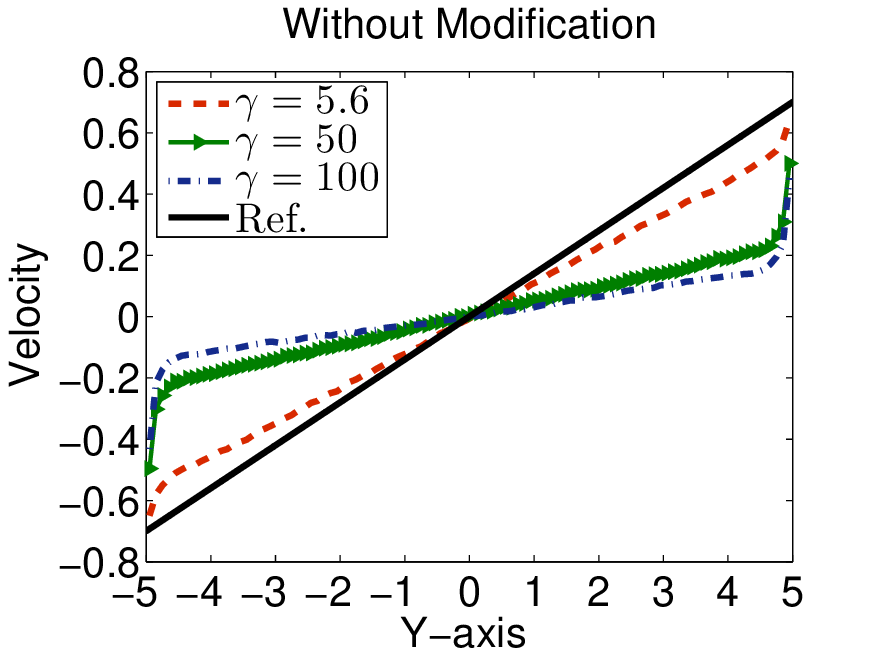}
\includegraphics[scale=0.5]{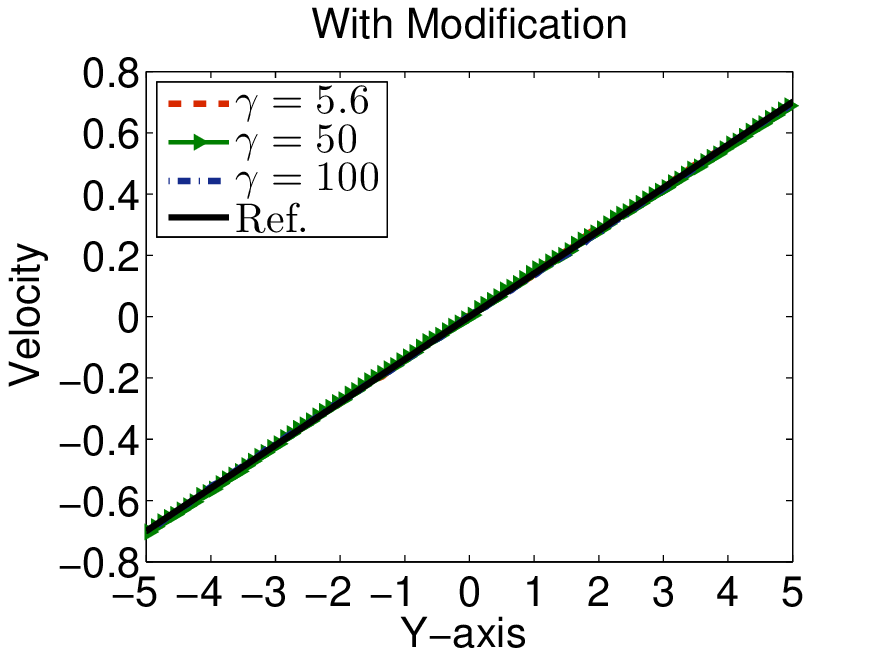}
\caption{\small Comparisons of the computed velocity profile in DPD with different values of the friction coefficient $\gamma$ under Lees--Edwards boundary conditions with shear rate $\dot{\gamma}=0.14$ ``without'' (left) and ``with'' (right) suitable modification in the relative velocity, respectively. The solid black line is the expected linear profile.}
\label{fig:PAdL_vProf_dpd_s1_modNO_vs_mod}
\end{figure}

A recent article of Chatterjee~\cite{Chatterjee2007} claimed that, owing to the dependence of inter-particle forces on the relative velocities of the particles, it is problematic to directly apply LEBC to DPD, especially in the high friction regime. As shown in Figure~\ref{fig:PAdL_vProf_dpd_s1_modNO_vs_mod} (left), where exactly the same setting as in the original article~\cite{Chatterjee2007} was used, as the friction increases, the velocity profile starts to (significantly) deviate away from the target linear profile. A simple remedy is to switch off the interactions of dissipative and random forces (i.e., the DPD thermostat) if one particle is within interacting range of an image of other particle near the boundaries where adjacent layers have different streaming velocities (i.e., the $y$-direction), as proposed in~\cite{Chatterjee2007}.

However, the finding of~\cite{Chatterjee2007} directly contradicts the principle of LEBC, which is translationally invariant and is intended to overcome the surface effects. In fact, as pointed in~\cite{Evans2008}, in no way can the particles actually sense the boundaries of any given box since the system is spatially homogeneous. Furthermore, our numerical experiments, which are in perfect \mbox{agreement} with theoretically expected behavior as shown in the right panel of Figure~\ref{fig:PAdL_vProf_dpd_s1_modNO_vs_mod} even in the high friction regime, suggest that LEBC might have been incorrectly implemented in~\cite{Chatterjee2007}. One possibility is that when calculating the relative velocity between one particle and an image of another, which is in a layer with different streaming velocity from its origin, the original velocity, rather than the ``modified velocity'' due to the different streaming velocities in different layers, was used as the velocity of the image particle. By neglecting the necessary modification, we obtained the left panel of Figure~\ref{fig:PAdL_vProf_dpd_s1_modNO_vs_mod}, while if the velocity of the image particle was properly modified the expected linear velocity profile was recovered in Figure~\ref{fig:PAdL_vProf_dpd_s1_modNO_vs_mod} (right) using otherwise exactly the same setting.

Overall, it should be emphasized that if one particle is interacting with the image of another under certain conditions, the relative velocity (in the $x$-direction) between them should be modified as follows:
\begin{align}
  & N_{\mathrm{L}} = 0 \label{eq:LEBC_Modified-1} \\
  &\mathrm{if} \, (\mathrm{fabs}(q^{y}_{ij}) > L_{y}/2) \quad N_{\mathrm{L}} = \mathrm{round} (q^{y}_{ij}/L_{y}) \label{eq:LEBC_Modified-2} \\
  & \hat{v}^{x}_{ij} = v^{x}_{ij} - N_{\mathrm{L}} \dot{\gamma}L_{y} \label{eq:LEBC_Modified-3}
\end{align}
where function ``$\mathrm{fabs}(\cdot)$'' returns the absolute value of the argument. Note that: (1) $q^{y}_{ij}$ in~\eqref{eq:LEBC_Modified-2} has to be evaluated either before the minimum image convention~\eqref{eq:LEBC_MIC-1}--\eqref{eq:LEBC_MIC-5} or by other proper ways to determine the actual distance between two interacting particles; (2) $v^{x}_{ij}$ in~\eqref{eq:LEBC_Modified-3} is the relative (``absolute'') velocity between the two particles and one has to take into account the effects of the streaming velocity as indicated.

We suspect that the necessary modification~\eqref{eq:LEBC_Modified-1}--\eqref{eq:LEBC_Modified-3} was not correctly implemented in~\cite{Chatterjee2007}, resulting in the nonphysical behavior as shown in Figure~\ref{fig:PAdL_vProf_dpd_s1_modNO_vs_mod} (left). It is not surprising that by switching off the DPD thermostat on interactions that cross certain boundaries would recover the expected linear velocity profile as shown in~\cite{Chatterjee2007}, since it directly avoids the situation described in~\eqref{eq:LEBC_Modified-1}--\eqref{eq:LEBC_Modified-3} where special attention has to be paid. Overall, the ``workaround'' does not provide any physical explanation, and could affect the underlying dynamics of the system, implying that it should be abandoned.

\subsubsection{Temperature in nonequilibrium molecular dynamics}

\begin{figure}[tb]
\centering
\includegraphics[scale=0.5]{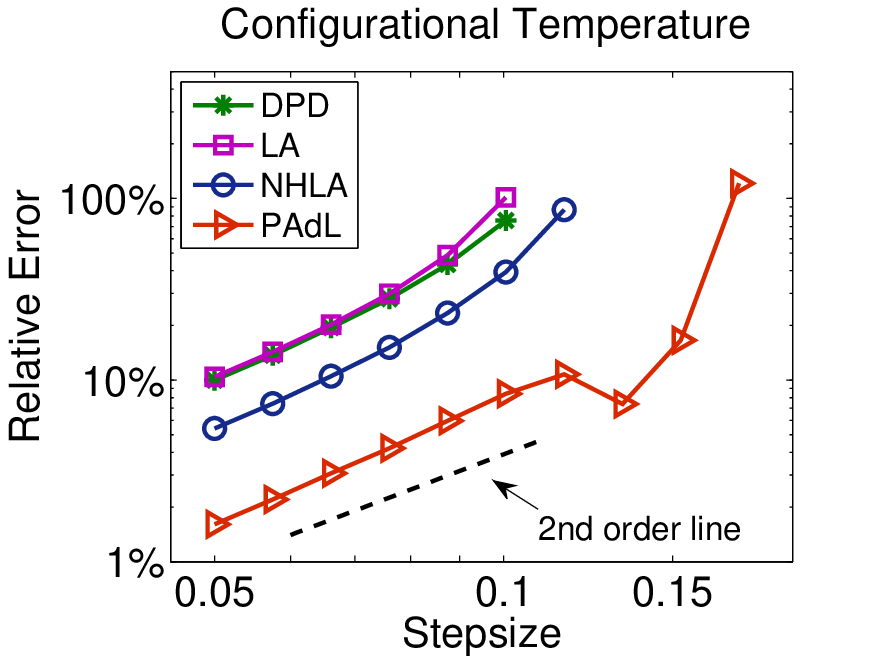}
\includegraphics[scale=0.5]{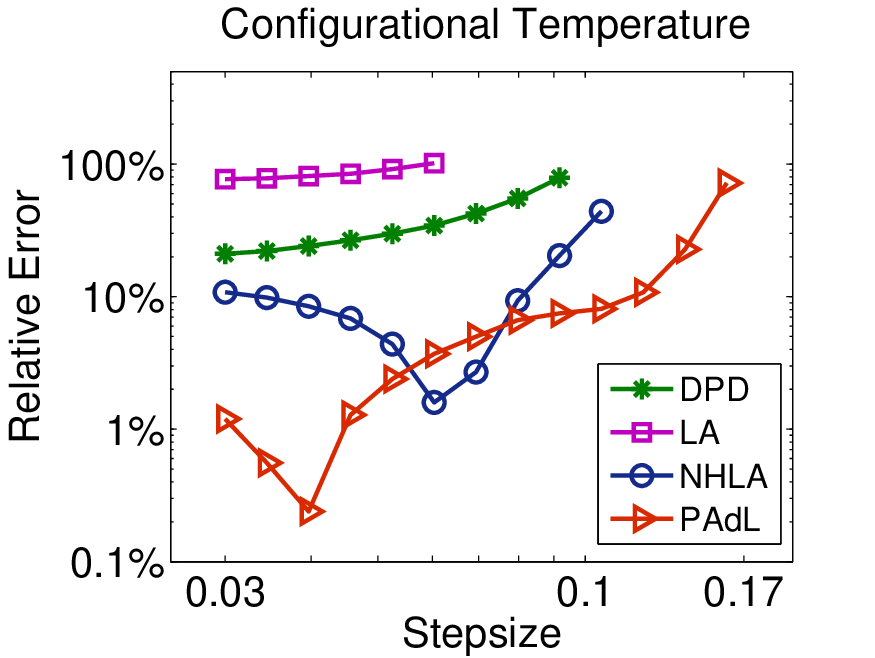}
\caption{\small Comparisons of the relative error in computed configurational temperature against stepsize by using various numerical methods with (effective) friction coefficient $\gamma=4.5$ under Lees--Edwards boundary conditions with shear rate $\dot{\gamma}=0.2$ (left) and $\dot{\gamma}=2$ (right). The format of the plots is the same as in Fig.~\ref{fig:PAdL_CT_U_Comp_gamma4d5}. }
\label{fig:PAdL_CT_Comp_gamma4d5_beta_d2_2}
\end{figure}

Another interesting question in NEMD is that ``What is the most appropriate way to measure the system temperature?'' Since the streaming velocity should be subtracted from the particle velocity before calculating the kinetic temperature, the normal definition should be modified as follows~\cite{Evans2008}:
\begin{equation}\label{eq:Kinetic_Temp_NEMD}
  k_{\mathrm{B}}\hat{T}_{\mathrm{k}} = \frac{1}{N_{\mathrm{d}}}\sum_{i} m_{i} (\mathbf{v}_{i}-\mathbf{u}) \cdot (\mathbf{v}_{i}-\mathbf{u}) \, ,
\end{equation}
where $N_{\mathrm{d}}$ is the number of degrees of freedom of the system and $\mathbf{u}$ denotes the corresponding streaming velocity at the location of particle $i$~\eqref{eq:LEBC_Streaming_V}. In other words, if $\mathbf{u}=0$, \eqref{eq:Kinetic_Temp_NEMD} reduces to the standard form.

If we can assume that the velocity profile is linear, as in uniform shear flow, we can just calculate and subsequently subtract it. However, as pointed in~\cite{Evans1986,Evans2008}, at higher shear rates and/or Reynolds numbers (i.e., the ratio of the inertial and viscous forces), the assumption of a linear streaming velocity profile is extremely dubious, even though Lees--Edwards boundary conditions are used. This issue was addressed in~\cite{Evans1986,Evans2008}, where the so-called PUT was proposed. The PUT allows the simulation itself define the local streaming velocity (see more details in~\cite{Evans1986,Evans2008,Whittle2010}). However, this is still not completely satisfactory since PUT assumes that the streaming velocity profile is stationary in time, whereas the profile could vary in time.

In DPD and related systems, temperature calculations can be based on relative velocities (e.g., the NHLA thermostat~\cite{Stoyanov2005}), which do not rely on the relationship between the absolute particle velocity and an underlying streaming velocity. As discussed in~\cite{Leimkuhler2015}, it is more desirable, especially in NEMD, to define the temperature solely based on the configurations, which leads to the configurational temperature defined in~\eqref{eq:Config_Temp} (see applications in~\cite{Delhommelle2001b,Delhommelle2001}). In addition, thermostats based on the configurational temperature have been widely used in NEMD with shear flows~\cite{Lue2002,Delhommelle2001a,Delhommelle2002}. Therefore, the configurational temperature formulation~\eqref{eq:Config_Temp} is used in our numerical experiments.

\begin{table}[tb]
\begin{center}
%\fontsize{9}{10.8}\selectfont{
\resizebox{1.0\textwidth}{!}{\begin{minipage}{\textwidth}
    \begin{tabular}{ |c|c|c|c|c| }
    \hline
    \textbf{Method} & \textbf{Critical stepsize} & \textbf{Maximal stepsize} &% \textbf{Force Calculation} &
    \textbf{CPU time} & \textbf{Scaled efficiency} \\ \hline
    DPD-VV      & 0.05 & 0.10 & %1 &
    20.212 & 100.0\% \\ \hline
    DPD-S1      & 0.05 & 0.11 & %1 &
    20.618 & 98.0\% \\ \hline
    DPD-Trotter & 0.05 & 0.11 & %1 &
    21.451 & 94.2\% \\ \hline
    Peters      & 0.05 & 0.11 & %1 &
    21.274 & 95.0\% \\ \hline
    LA          & 0.05 & 0.10 & %1 &
    18.048 & 112.0\% \\ \hline
    NHLA        & 0.07 & 0.13 & %1 &
    18.691 & 151.4\% \\ \hline
%    PNH         & 0.05 & 0.08 & 1 & 16.910 & 119.5\% \\ \hline
%    PNHL-S      & 0.08 & 0.17 & 1 & 21.589 & 149.8\% \\ \hline
%    PNHL-N      & 0.17 & 0.23 & 2 & 38.421 & 178.9\% \\ \hline
    PAdL        & 0.13 & 0.17 & %1 &
     23.103 & 227.5\% \\
    \hline
    \end{tabular}
\caption[Table caption text]{\small Comparisons of the computational efficiency of the various numerical methods in the moderate (effective) friction regime of $\gamma=4.5$ under Lees--Edwards boundary conditions with shear rate $\dot{\gamma}=0.2$. The format of the table is the same as in Table~\ref{table:efficiency_eq_gamma_4d5}.}
\label{table:efficiency_noneq}
\end{minipage} } %}
\end{center}
\end{table}

\subsubsection{Results}

Figure~\ref{fig:PAdL_CT_Comp_gamma4d5_beta_d2_2} compares the configurational temperature control of various systems described under LEBC with different shear rates. As can be shown from the figure, when the shear rate is relatively small ($\dot{\gamma}=0.2$, which is larger than that of Figure~\ref{fig:PAdL_vProf_dpd_s1_modNO_vs_mod}), the behavior is largely similar to that of Figure~\ref{fig:PAdL_CT_U_Comp_gamma4d5} (left): all the methods appear to show second order convergence to the invariant measure, and the newly proposed PAdL method achieves one order of magnitude improvement in numerical accuracy over both DPD and LA methods, both of which are slightly worse than the NHLA method. The overall numerical efficiency was compared in Table~\ref{table:efficiency_noneq}. Again the PAdL method is by far the most efficient method of all, which has an about 130\% improvement over the benchmark DPD-VV method.

When the shear rate is relatively high ($\dot{\gamma}=2$), as shown in Figure~\ref{fig:PAdL_CT_Comp_gamma4d5_beta_d2_2} (right), all the methods appear to lose the clear second order convergence previously observed. The LA thermostat, displaying  large relative error even when the stepsize is relatively small, appears to be most vulnerable to the high shear rate, with the DPD method being slightly better. While exhibiting some unexpected behavior, both NHLA and PAdL have better numerical accuracy control than the other two. It can be seen that, as the stepsize increases, the relative error of the PAdL method starts to decrease before growing as would be expected. We believe this unexpected decrease at the beginning is due to sampling errors, which is more likely to be observed in the high accuracy regime, since increasing the system size (i.e., number of particles) would resolve this issue. Nevertheless, the PAdL method consistently achieves an order of magnitude improvement in numerical accuracy over the DPD method.

\section{Conclusions}
\label{sec:Conclusions}

The most significant contributions of this article are as follows: (i) the introduction of a new negative-feedback-loop-controlled formulation of DPD, i.e., the PAdL thermostat, that preserves the dynamical behavior of DPD while enhancing the accuracy of averages, and (ii) an analysis explaining the second order convergence of the nonsymmetric PNHL method~\cite{Leimkuhler2015} for a broad class of observables.  We have, additionally, clarified the treatment of Lees--Edwards boundary conditions for modeling shear flow.     PAdL (and PNHL, in the low friction regime) effectively double the performance of DPD, are easy to implement, and rest on a solid foundation of theory and previous numerical experience in the Langevin dynamics context.

PAdL and PNHL use auxiliary dynamical variables (like the ``global demon'' of Nos\'{e}--Hoover dynamics), whose interaction with the physical variables is controlled by a coupling parameter. In the deterministic case, the choice of the coupling coefficient (or ``thermal mass'') can be an added complication. However, our previous experience with the Nos\'{e}--Hoover--Langevin method (as well as in the numerical experiments of this article) has demonstrated that, once stochastic noise is present, the character of the parameter dramatically changes, so that the method properties are robust to a much wider range of values.

The key idea that is exploited here in the formulation of PAdL is that the thermostat can correct for numerical errors.  This builds on the observation of~\cite{Sivak2013} that the discretization error can be interpreted as irreversible work that induces effective heating.  What is less obvious is that the negative feedback loop controller used here, which introduces additional dynamics that also must be discretized, does not distort the invariant distribution, but we surmise that this is a consequence of the simple form of this equation and its simple discretization which maintains the form of a discrete control law.

The usefulness of the PAdL scheme will be felt most strongly in simulations that have a strong hydrodynamic character, e.g., nonequilibrium systems undergoing shear flow.   In other cases, for example, models in the solid or gas states, one would expect that Langevin dynamics or the PNHL method described here will be of more value.  A secondary benefit of the PAdL thermostat which we have not explored in this article, is its ability to correct thermodynamics automatically for errors through the computation of the conservative force.

DPD and PAdL would likely have similar ergodic properties which depend on the choice of weight functions and potentials.  It will be of interest to explore this issue further in future studies.  These choices will also influence the accuracy and stability of the numerical methods and thus the potential efficiency improvements available.  Exploration is needed of the numerical behavior when the potential energies arise from tabled or interpolated data.

It should be noted that the idea of ``transverse'' DPD~\cite{Junghans2008,Li2014a} (i.e., including the transverse component of dissipative and random forces) could easily be incorporated into PAdL. The parallelization of the proposed method of PAdL is not completely trivial, but is similar to the task of parallelizing Shardlow-like schemes. The problem has recently been addressed by Larentzos et al.~\cite{Larentzos2014}.

\section*{Acknowledgements}

The authors thank Michael Allen and anonymous referees for valuable suggestions and comments. The authors acknowledge the support of the Engineering and Physical Sciences Research Council (UK) through the project of SI2-CHE: ExTASY: Extensible Tools for Advanced Sampling and analYsis (Grant No. EP/K039512/1).


%\appendix

\begin{appendices}
  \renewcommand\thetable{\thesection\arabic{table}}
  \renewcommand\thefigure{\thesection\arabic{figure}}

\section{Integration schemes}
\label{sec:Appendix_Schemes}

We list detailed integration steps for each method described in the article here. Verlet neighbor lists~\cite{Verlet1967} are used throughout each method.

\subsection*{Symmetric pairwise Nos\'{e}--Hoover--Langevin thermostat: \mbox{PNHL-S}}

For each particle $i$,
\begin{align*}
  \mathbf{q}_{i}^{n+1/2} &=  \mathbf{q}_{i}^{n} + hm_{i}^{-1}\mathbf{p}_{i}^{n}/2 \, , \\
  \mathbf{p}_{i}^{n+1/4} &= \mathbf{p}_{i}^{n} + h\mathbf{F}_{i}^{\mathrm{C}}(\mathbf{q}^{n+1/2})/2 \, ,
\end{align*}
where $\mathbf{F}^{\mathrm{C}}_{i}(\mathbf{q}) = -\nabla_{\mathbf{q}_{i}}U(\mathbf{q})$ are the total conservative forces acting on particle $i$.

\noindent For each interacting pair within cutoff radius ($r_{ij}<r_{\mathrm{c}}$),
\begin{align*}
  \mathbf{p}^{n+2/4}_{i} &= \mathbf{p}^{n+1/4}_{i} + m_{ij}\Delta v_{ij}(\mathbf{q}^{n+1/2},\mathbf{p}^{n+1/4},\xi^{n}) \hat{\mathbf{q}}^{n+1/2}_{ij}/2 \, , \\
  \mathbf{p}^{n+2/4}_{j} &= \mathbf{p}^{n+1/4}_{j} - m_{ij}\Delta v_{ij}(\mathbf{q}^{n+1/2},\mathbf{p}^{n+1/4},\xi^{n}) \hat{\mathbf{q}}^{n+1/2}_{ij}/2 \, ,
\end{align*}
where $m_{ij}=m_{i}m_{j}/(m_{i}+m_{j})$ and
\begin{equation*}
    \Delta v_{ij} = \left( \hat{\mathbf{q}}_{ij} \cdot \mathbf{v}_{ij} \right) \left( \exp\left(-\xi \omega^{\mathrm{D}}(r_{ij}) (h/2) / m_{ij} \right) - 1 \right) \, .
\end{equation*}

\noindent For additional variable $\xi$,
\begin{align*}
  \xi^{n+1/3} &=  \xi^{n} + hG(\mathbf{q}^{n+1/2},\mathbf{p}^{n+2/4})/2 \, , \\
  \xi^{n+2/3} &= e^{-\tilde{\gamma} h}\xi^{n+1/3} + \sqrt{ k_{\mathrm{B}}T(1-e^{-2\tilde{\gamma} h})/\mu }\mathrm{R} \, , \\
  \xi^{n+1} &=  \xi^{n+2/3} + hG(\mathbf{q}^{n+1/2},\mathbf{p}^{n+2/4})/2 \, ,
\end{align*}
where
\begin{equation*}
    G(\mathbf{q},\mathbf{p}) = {\mu}^{-1}\sum_{i}\sum_{j>i}\omega^{\mathrm{D}}(r_{ij}) \left[ \left( \mathbf{v}_{ij}\cdot\hat{\mathbf{q}}_{ij} \right)^{2} - k_{\mathrm{B}}T/m_{ij}  \right]
\end{equation*}
and $\mathrm{R}$ are normally distributed variables with zero mean and unit variance.

\noindent For each interacting pair within cutoff radius ($r_{ij}<r_{\mathrm{c}}$),
\begin{align*}
  \mathbf{p}^{n+3/4}_{i} &= \mathbf{p}^{n+2/4}_{i} + m_{ij}\Delta v_{ij}(\mathbf{q}^{n+1/2},\mathbf{p}^{n+2/4},\xi^{n+1}) \hat{\mathbf{q}}^{n+1/2}_{ij}/2 \, , \\
  \mathbf{p}^{n+3/4}_{j} &= \mathbf{p}^{n+2/4}_{j} - m_{ij}\Delta v_{ij}(\mathbf{q}^{n+1/2},\mathbf{p}^{n+2/4},\xi^{n+1}) \hat{\mathbf{q}}^{n+1/2}_{ij}/2 \, .
\end{align*}

\noindent For each particle $i$,
\begin{align*}
  \mathbf{p}_{i}^{n+1} &= \mathbf{p}_{i}^{n+3/4} + h\mathbf{F}_{i}^{\mathrm{C}}(\mathbf{q}^{n+1/2})/2 \, , \\
  \mathbf{q}_{i}^{n+1} &=  \mathbf{q}_{i}^{n+1/2} + hm_{i}^{-1}\mathbf{p}_{i}^{n+1}/2 \, .
\end{align*}

\subsection*{Nonsymmetric pairwise Nos\'{e}--Hoover--Langevin thermostat: \mbox{PNHL-N}}

For each particle $i$,
\begin{align*}
  \mathbf{q}_{i}^{n+1/2} &=  \mathbf{q}_{i}^{n} + hm_{i}^{-1}\mathbf{p}_{i}^{n}/2 \, , \\
  \mathbf{p}_{i}^{n+1/4} &= \mathbf{p}_{i}^{n} + h\mathbf{F}_{i}^{\mathrm{C}}(\mathbf{q}^{n+1/2})/2 \, .
\end{align*}

\noindent For each interacting pair within cutoff radius ($r_{ij}<r_{\mathrm{c}}$),
\begin{align*}
  \mathbf{p}^{n+2/4}_{i} &= \mathbf{p}^{n+1/4}_{i} + m_{ij}\Delta v_{ij}(\mathbf{q}^{n+1/2},\mathbf{p}^{n+1/4},\xi^{n}) \hat{\mathbf{q}}^{n+1/2}_{ij}/2 \, , \\
  \mathbf{p}^{n+2/4}_{j} &= \mathbf{p}^{n+1/4}_{j} - m_{ij}\Delta v_{ij}(\mathbf{q}^{n+1/2},\mathbf{p}^{n+1/4},\xi^{n}) \hat{\mathbf{q}}^{n+1/2}_{ij}/2 \, ,
\end{align*}
where
\begin{equation*}
    \Delta v_{ij} = \left( \hat{\mathbf{q}}_{ij} \cdot \mathbf{v}_{ij} \right) \left( \exp\left(-\xi \omega^{\mathrm{D}}(r_{ij}) (h/2) / m_{ij} \right) - 1 \right) \, .
\end{equation*}

\noindent For additional variable $\xi$,
\begin{align*}
  \xi^{n+1/3} &=  \xi^{n} + hG(\mathbf{q}^{n+1/2},\mathbf{p}^{n+2/4})/2 \, , \\
  \xi^{n+2/3} &= e^{-\tilde{\gamma} h}\xi^{n+1/3} + \sqrt{ k_{\mathrm{B}}T(1-e^{-2\tilde{\gamma} h})/\mu }\mathrm{R} \, , \\
  \xi^{n+1} &=  \xi^{n+2/3} + hG(\mathbf{q}^{n+1/2},\mathbf{p}^{n+2/4})/2 \, .
\end{align*}

\noindent For each interacting pair within cutoff radius ($r_{ij}<r_{\mathrm{c}}$),
\begin{align*}
  \mathbf{p}^{n+3/4}_{i} &= \mathbf{p}^{n+2/4}_{i} + m_{ij}\Delta v_{ij}(\mathbf{q}^{n+1/2},\mathbf{p}^{n+2/4},\xi^{n+1}) \hat{\mathbf{q}}^{n+1/2}_{ij}/2 \, , \\
  \mathbf{p}^{n+3/4}_{j} &= \mathbf{p}^{n+2/4}_{j} - m_{ij}\Delta v_{ij}(\mathbf{q}^{n+1/2},\mathbf{p}^{n+2/4},\xi^{n+1}) \hat{\mathbf{q}}^{n+1/2}_{ij}/2 \, .
\end{align*}

\noindent For each particle $i$,
\begin{align*}
  \mathbf{q}_{i}^{n+1} &=  \mathbf{q}_{i}^{n+1/2} + hm_{i}^{-1}\mathbf{p}_{i}^{n+3/4}/2 \, , \\
  \mathbf{p}_{i}^{n+1} &= \mathbf{p}_{i}^{n+3/4} + h\mathbf{F}_{i}^{\mathrm{C}}(\mathbf{q}^{n+1})/2 \, .
\end{align*}

\subsection*{Pairwise adaptive Langevin thermostat: PAdL}

For each particle $i$,
\begin{align*}
  \mathbf{q}_{i}^{n+1/2} &=  \mathbf{q}_{i}^{n} + hm_{i}^{-1}\mathbf{p}_{i}^{n}/2 \, , \\
  \mathbf{p}_{i}^{n+1/4} &= \mathbf{p}_{i}^{n} + h\mathbf{F}_{i}^{\mathrm{C}}(\mathbf{q}^{n+1/2})/2 \, .
\end{align*}
\noindent For each interacting pair within cutoff radius ($r_{ij}<r_{\mathrm{c}}$),
\begin{align*}
  \mathbf{p}^{n+2/4}_{i} &= \mathbf{p}^{n+1/4}_{i} + m_{ij}\Delta v_{ij}(\mathbf{q}^{n+1/2},\mathbf{p}^{n+1/4},\xi^{n}) \hat{\mathbf{q}}^{n+1/2}_{ij} \, , \\
  \mathbf{p}^{n+2/4}_{j} &= \mathbf{p}^{n+1/4}_{j} - m_{ij}\Delta v_{ij}(\mathbf{q}^{n+1/2},\mathbf{p}^{n+1/4},\xi^{n}) \hat{\mathbf{q}}^{n+1/2}_{ij} \, ,
\end{align*}
where, \\
\indent if ($\xi^{n} \neq 0$):
\begin{equation*}
    \Delta v_{ij} = \left( \hat{\mathbf{q}}_{ij} \cdot \mathbf{v}_{ij} \right) \left( \exp(-\tilde{\tau} h/2) - 1 \right) + \sigma \sqrt{ \frac{1-\exp(-\tilde{\tau} h)}{2\xi^{n}m_{ij}} } \mathrm{R}_{ij} \, ,
\end{equation*}
where $\tilde{\tau}=\xi \omega^{\mathrm{D}}(r_{ij}) / m_{ij}$ and $\mathrm{R}_{ij}$ are normally distributed variables with zero mean and unit variance;

else:
\begin{equation*}
    \Delta v_{ij} =  \sigma \left(\omega^{\mathrm{R}}(r_{ij})/m_{ij}\right) \sqrt{h/2} \mathrm{R}_{ij} \, .
\end{equation*}

\noindent For additional variable $\xi$,
\begin{equation*}
  \xi^{n+1} =  \xi^{n} + hG(\mathbf{q}^{n+1/2},\mathbf{p}^{n+2/4})/2 \, .
\end{equation*}

\noindent For each interacting pair within cutoff radius ($r_{ij}<r_{\mathrm{c}}$),
\begin{align*}
  \mathbf{p}^{n+3/4}_{i} &= \mathbf{p}^{n+2/4}_{i} + m_{ij}\Delta v_{ij}(\mathbf{q}^{n+1/2},\mathbf{p}^{n+2/4},\xi^{n+1}) \hat{\mathbf{q}}^{n+1/2}_{ij} \, , \\
  \mathbf{p}^{n+3/4}_{j} &= \mathbf{p}^{n+2/4}_{j} - m_{ij}\Delta v_{ij}(\mathbf{q}^{n+1/2},\mathbf{p}^{n+2/4},\xi^{n+1}) \hat{\mathbf{q}}^{n+1/2}_{ij} \, ,
\end{align*}
where, \\
\indent if ($\xi^{n+1} \neq 0$):
\begin{equation*}
    \Delta v_{ij} = \left( \hat{\mathbf{q}}_{ij} \cdot \mathbf{v}_{ij} \right) \left( \exp(-\tilde{\tau} h/2) - 1 \right) + \sigma \sqrt{ \frac{1-\exp(-\tilde{\tau} h)}{2\xi^{n+1}m_{ij}} } \mathrm{R}_{ij} \, .
\end{equation*}

else:
\begin{equation*}
    \Delta v_{ij} =  \sigma \left(\omega^{\mathrm{R}}(r_{ij})/m_{ij}\right) \sqrt{h/2} \mathrm{R}_{ij} \, .
\end{equation*}

\noindent For each particle $i$,
\begin{align*}
  \mathbf{p}_{i}^{n+1} &= \mathbf{p}_{i}^{n+3/4} + h\mathbf{F}_{i}^{\mathrm{C}}(\mathbf{q}^{n+1/2})/2 \, , \\
  \mathbf{q}_{i}^{n+1} &=  \mathbf{q}_{i}^{n+1/2} + hm_{i}^{-1}\mathbf{p}_{i}^{n+1}/2 \, .
\end{align*}

\end{appendices}

\bibliographystyle{is-abbrv}
%\bibliographystyle{unsrt}

%\bibliography{sample}
\bibliography{refs}

\begin{thebibliography}{10}

\bibitem{Abdulle2014a}
A.~Abdulle, G.~Vilmart, and K.~C. Zygalakis.
\newblock High order numerical approximation of the invariant measure of
  ergodic {SDE}s.
\newblock {\em SIAM Journal on Numerical Analysis}, 52\penalty0 (4):\penalty0
  1600--1622, 2014.

\bibitem{Abdulle2014}
A.~Abdulle, G.~Vilmart, and K.~C. Zygalakis.
\newblock Long time accuracy of {L}ie--{T}rotter splitting methods for
  {L}angevin dynamics.
\newblock {\em SIAM Journal on Numerical Analysis}, 53\penalty0 (1):\penalty0
  1--16, 2015.

\bibitem{Allen2006}
M.~P. Allen.
\newblock Configurational temperature in membrane simulations using dissipative
  particle dynamics.
\newblock {\em The Journal of Physical Chemistry B}, 110\penalty0 (8):\penalty0
  3823--3830, 2006.

\bibitem{Allen2007}
M.~P. Allen and F.~Schmid.
\newblock A thermostat for molecular dynamics of complex fluids.
\newblock {\em Molecular Simulation}, 33\penalty0 (1-2):\penalty0 21--26, 2007.

\bibitem{Allen1989}
M.~P. Allen and D.~J. Tildesley.
\newblock {\em Computer Simulation of Liquids}.
\newblock Oxford University Press, 1989.

\bibitem{Ashurst1975}
W.~T. Ashurst and W.~G. Hoover.
\newblock Dense-fluid shear viscosity via nonequilibrium molecular dynamics.
\newblock {\em Physical Review A}, 11\penalty0 (2):\penalty0 658--678, 1975.

\bibitem{Backer2005}
J.~A. Backer, C.~P. Lowe, H.~C.~J. Hoefsloot, and P.~D. Iedema.
\newblock Poiseuille flow to measure the viscosity of particle model fluids.
\newblock {\em The Journal of Chemical Physics}, 122:\penalty0 154503, 2005.

\bibitem{Besold2000}
G.~Besold, I.~Vattulainen, M.~Karttunen, and J.~M. Polson.
\newblock Towards better integrators for dissipative particle dynamics
  simulations.
\newblock {\em Physical Review E}, 62\penalty0 (6):\penalty0 R7611, 2000.

\bibitem{Boek1996}
E.~S. Boek, P.~V. Coveney, and H.~N.~W. Lekkerkerker.
\newblock Computer simulation of rheological phenomena in dense colloidal
  suspensions with dissipative particle dynamics.
\newblock {\em Journal of Physics: Condensed Matter}, 8\penalty0 (47):\penalty0
  9509--9512, 1996.

\bibitem{Boek1997}
E.~S. Boek, P.~V. Coveney, H.~N.~W. Lekkerkerker, and P.~van~der Schoot.
\newblock Simulating the rheology of dense colloidal suspensions using
  dissipative particle dynamics.
\newblock {\em Physical Review E}, 55\penalty0 (3):\penalty0 3124, 1997.

\bibitem{Bou-Rabee2010}
N.~Bou-Rabee and H.~Owhadi.
\newblock Long-run accuracy of variational integrators in the stochastic
  context.
\newblock {\em SIAM Journal on Numerical Analysis}, 48\penalty0 (1):\penalty0
  278--297, 2010.

\bibitem{Braga2005}
C.~Braga and K.~P. Travis.
\newblock A configurational temperature {N}os{\'e}--{H}oover thermostat.
\newblock {\em The Journal of Chemical Physics}, 123:\penalty0 134101, 2005.

\bibitem{Chatterjee2007}
A.~Chatterjee.
\newblock Modification to {L}ees--{E}dwards periodic boundary condition for
  dissipative particle dynamics simulation with high dissipation rates.
\newblock {\em Molecular Simulation}, 33\penalty0 (15):\penalty0 1233--1236,
  2007.

\bibitem{Chaudhri2010}
A.~Chaudhri and J.~R. Lukes.
\newblock Velocity and stress autocorrelation decay in isothermal dissipative
  particle dynamics.
\newblock {\em Physical Review E}, 81\penalty0 (2):\penalty0 026707, 2010.

\bibitem{Ciccotti1980}
G.~Ciccotti and A.~Tenenbaum.
\newblock Canonical ensemble and nonequilibrium states by molecular dynamics.
\newblock {\em Journal of Statistical Physics}, 23\penalty0 (6):\penalty0
  767--772, 1980.

\bibitem{DeFabritiis2006}
G.~De~Fabritiis, M.~Serrano, P.~Espa{\~n}ol, and P.~Coveney.
\newblock Efficient numerical integrators for stochastic models.
\newblock {\em Physica A: Statistical Mechanics and its Applications},
  361\penalty0 (2):\penalty0 429--440, 2006.

\bibitem{Debussche2012}
A.~Debussche and E.~Faou.
\newblock Weak backward error analysis for {SDE}s.
\newblock {\em SIAM Journal on Numerical Analysis}, 50\penalty0 (3):\penalty0
  1735--1752, 2012.

\bibitem{Delhommelle2001b}
J.~Delhommelle and D.~J. Evans.
\newblock Configurational temperature profile in confined fluids. {I}. {A}tomic
  fluid.
\newblock {\em The Journal of Chemical Physics}, 114\penalty0 (14):\penalty0
  6229, 2001.

\bibitem{Delhommelle2001}
J.~Delhommelle and D.~J. Evans.
\newblock Configurational temperature profile in confined fluids. {II}.
  {M}olecular fluids.
\newblock {\em The Journal of Chemical Physics}, 114\penalty0 (14):\penalty0
  6236, 2001.

\bibitem{Delhommelle2001a}
J.~Delhommelle and D.~J. Evans.
\newblock Configurational temperature thermostat for fluids undergoing shear
  flow: application to liquid chlorine.
\newblock {\em Molecular Physics}, 99\penalty0 (21):\penalty0 1825--1829, 2001.

\bibitem{Delhommelle2002}
J.~Delhommelle and D.~J. Evans.
\newblock Correspondence between configurational temperature and molecular
  kinetic temperature thermostats.
\newblock {\em The Journal of Chemical Physics}, 117\penalty0 (13):\penalty0
  6016, 2002.

\bibitem{Duenweg1993}
B.~D{\"u}nweg.
\newblock Molecular dynamics algorithms and hydrodynamic screening.
\newblock {\em The Journal of Chemical Physics}, 99\penalty0 (9):\penalty0
  6977--6982, 1993.

\bibitem{Espanol1995a}
P.~Espa{\~n}ol.
\newblock Hydrodynamics from dissipative particle dynamics.
\newblock {\em Physical Review E}, 52:\penalty0 1734--1742, 1995.

\bibitem{Espanol1995}
P.~Espa{\~n}ol and P.~Warren.
\newblock Statistical mechanics of dissipative particle dynamics.
\newblock {\em Europhysics Letters}, 30\penalty0 (4):\penalty0 191, 1995.

\bibitem{Evans2008}
D.~J. Evans and G.~Morriss.
\newblock {\em Statistical Mechanics of Nonequilibrium Liquids}.
\newblock Cambridge University Press, 2008.

\bibitem{Evans1986}
D.~J. Evans and G.~P. Morriss.
\newblock Shear thickening and turbulence in simple fluids.
\newblock {\em Physical Review Letters}, 56\penalty0 (20):\penalty0 2172, 1986.

\bibitem{Farago2016}
O.~Farago and N.~Gr{\o}nbech-Jensen.
\newblock On the connection between dissipative particle dynamics and the
  {I}t\^{o}--{S}tratonovich dilemma.
\newblock {\em The Journal of Chemical Physics}, 144\penalty0 (8):\penalty0
  084102, 2016.

\bibitem{Fedosov2010}
D.~A. Fedosov, G.~E. Karniadakis, and B.~Caswell.
\newblock Steady shear rheometry of dissipative particle dynamics models of
  polymer fluids in reverse {P}oiseuille flow.
\newblock {\em The Journal of Chemical Physics}, 132\penalty0 (14):\penalty0
  144103, 2010.

\bibitem{Fedosov2011}
D.~A. Fedosov, W.~Pan, B.~Caswell, G.~Gompper, and G.~E. Karniadakis.
\newblock Predicting human blood viscosity in silico.
\newblock {\em Proceedings of the National Academy of Sciences}, 108\penalty0
  (29):\penalty0 11772--11777, 2011.

\bibitem{Frenkel2001}
D.~Frenkel and B.~Smit.
\newblock {\em Understanding Molecular Simulation: From Algorithms to
  Applications, Second Edition}.
\newblock Academic Press, 2001.

\bibitem{Frisch1986}
U.~Frisch, B.~Hasslacher, and Y.~Pomeau.
\newblock Lattice-gas automata for the {N}avier--{S}tokes equation.
\newblock {\em Physical Review Letters}, 56\penalty0 (14):\penalty0 1505--1508,
  1986.

\bibitem{Givon2004}
D.~Givon, R.~Kupferman, and A.~Stuart.
\newblock Extracting macroscopic dynamics: model problems and algorithms.
\newblock {\em Nonlinearity}, 17\penalty0 (6):\penalty0 R55--R127, 2004.

\bibitem{Groot1997}
R.~D. Groot and P.~B. Warren.
\newblock Dissipative particle dynamics: Bridging the gap between atomistic and
  mesoscopic simulation.
\newblock {\em The Journal of Chemical Physics}, 107:\penalty0 4423, 1997.

\bibitem{Hansen2006}
J.-P. Hansen and I.~McDonald.
\newblock {\em Theory of Simple Liquids, Third Edition}.
\newblock Academic Press, 2006.

\bibitem{Hoogerbrugge1992}
P.~Hoogerbrugge and J.~Koelman.
\newblock Simulating microscopic hydrodynamic phenomena with dissipative
  particle dynamics.
\newblock {\em Europhysics Letters}, 19\penalty0 (3):\penalty0 155, 1992.

\bibitem{Hoover1991}
W.~G. Hoover.
\newblock {\em Computational Statistical Mechanics, Studies in Modern
  Thermodynamics}.
\newblock Elsevier Science, 1991.

\bibitem{Jakobsen2005}
A.~F. Jakobsen, O.~G. Mouritsen, and G.~Besold.
\newblock Artifacts in dynamical simulations of coarse-grained model lipid
  bilayers.
\newblock {\em The Journal of Chemical Physics}, 122\penalty0 (20):\penalty0
  204901, 2005.

\bibitem{Jones2011}
A.~Jones and B.~Leimkuhler.
\newblock Adaptive stochastic methods for sampling driven molecular systems.
\newblock {\em The Journal of Chemical Physics}, 135\penalty0 (8):\penalty0
  084125, 2011.

\bibitem{Joubaud2012}
R.~Joubaud and G.~Stoltz.
\newblock Nonequilibrium shear viscosity computations with {L}angevin dynamics.
\newblock {\em Multiscale Modeling \& Simulation}, 10\penalty0 (1):\penalty0
  191--216, 2012.

\bibitem{Junghans2008}
C.~Junghans, M.~Praprotnik, and K.~Kremer.
\newblock Transport properties controlled by a thermostat: An extended
  dissipative particle dynamics thermostat.
\newblock {\em Soft Matter}, 4\penalty0 (1):\penalty0 156--161, 2008.

\bibitem{Kloeden1992}
P.~E. Kloeden and E.~Platen.
\newblock {\em Numerical Solution of Stochastic Differential Equations}.
\newblock Springer, 1992.

\bibitem{Larentzos2014}
J.~P. Larentzos, J.~K. Brennan, J.~D. Moore, M.~L{\'i}sal, and W.~D. Mattson.
\newblock Parallel implementation of isothermal and isoenergetic dissipative
  particle dynamics using {S}hardlow-like splitting algorithms.
\newblock {\em Computer Physics Communications}, 185\penalty0 (7):\penalty0
  1987--1998, 2014.

\bibitem{Lees1972}
A.~Lees and S.~Edwards.
\newblock The computer study of transport processes under extreme conditions.
\newblock {\em Journal of Physics C: Solid State Physics}, 5\penalty0
  (15):\penalty0 1921, 1972.

\bibitem{Lei2010}
H.~Lei, B.~Caswell, and G.~E. Karniadakis.
\newblock Direct construction of mesoscopic models from microscopic
  simulations.
\newblock {\em Physical Review E}, 81\penalty0 (2):\penalty0 026704, 2010.

\bibitem{Leimkuhler2013}
B.~Leimkuhler and C.~Matthews.
\newblock Rational construction of stochastic numerical methods for molecular
  sampling.
\newblock {\em Applied Mathematics Research eXpress}, 2013\penalty0
  (1):\penalty0 34--56, 2013.

\bibitem{Leimkuhler2013a}
B.~Leimkuhler and C.~Matthews.
\newblock Robust and efficient configurational molecular sampling via
  {L}angevin dynamics.
\newblock {\em The Journal of Chemical Physics}, 138:\penalty0 174102, 2013.

\bibitem{Leimkuhler2015b}
B.~Leimkuhler and C.~Matthews.
\newblock {\em Molecular Dynamics: With Deterministic and Stochastic Numerical
  Methods}.
\newblock Springer, 2015.

\bibitem{Leimkuhler2013c}
B.~Leimkuhler, C.~Matthews, and G.~Stoltz.
\newblock The computation of averages from equilibrium and nonequilibrium
  {L}angevin molecular dynamics.
\newblock {\em IMA Journal of Numerical Analysis}, 36\penalty0 (1):\penalty0
  13--79, 2016.

\bibitem{Leimkuhler2009}
B.~Leimkuhler, E.~Noorizadeh, and F.~Theil.
\newblock A gentle stochastic thermostat for molecular dynamics.
\newblock {\em Journal of Statistical Physics}, 135\penalty0 (2):\penalty0
  261--277, 2009.

\bibitem{Leimkuhler2015}
B.~Leimkuhler and X.~Shang.
\newblock On the numerical treatment of dissipative particle dynamics and
  related systems.
\newblock {\em Journal of Computational Physics}, 280:\penalty0 72--95, 2015.

\bibitem{Leimkuhler2015a}
B.~Leimkuhler and X.~Shang.
\newblock Adaptive thermostats for noisy gradient systems.
\newblock {\em SIAM Journal on Scientific Computing}, 38\penalty0 (2):\penalty0
  A712--A736, 2016.

\bibitem{Li2013}
X.~Li, P.~M. Vlahovska, and G.~E. Karniadakis.
\newblock Continuum- and particle-based modeling of shapes and dynamics of red
  blood cells in health and disease.
\newblock {\em Soft Matter}, 9\penalty0 (1):\penalty0 28--37, 2013.

\bibitem{Li2014a}
Z.~Li, X.~Bian, B.~Caswell, and G.~E. Karniadakis.
\newblock Construction of dissipative particle dynamics models for complex
  fluids via the {M}ori--{Z}wanzig formulation.
\newblock {\em Soft Matter}, 10\penalty0 (43):\penalty0 8659--8672, 2014.

\bibitem{Li2015b}
Z.~Li, J.~R. Kermode, and A.~De~Vita.
\newblock Molecular dynamics with on-the-fly machine learning of
  quantum-mechanical forces.
\newblock {\em Physical Review Letters}, 114\penalty0 (9):\penalty0 096405,
  2015.

\bibitem{Lowe1999}
C.~Lowe.
\newblock An alternative approach to dissipative particle dynamics.
\newblock {\em Europhysics Letters}, 47\penalty0 (2):\penalty0 145, 1999.

\bibitem{Lue2002}
L.~Lue, O.~G. Jepps, J.~Delhommelle, and D.~J. Evans.
\newblock Configurational thermostats for molecular systems.
\newblock {\em Molecular Physics}, 100\penalty0 (14):\penalty0 2387--2395,
  2002.

\bibitem{Lyubartsev1995}
A.~Lyubartsev and A.~Laaksonen.
\newblock Calculation of effective interaction potentials from radial
  distribution functions: A reverse {M}onte {C}arlo approach.
\newblock {\em Physical Review E}, 52\penalty0 (4):\penalty0 3730--3737, 1995.

\bibitem{Moeendarbary2009}
E.~Moeendarbary, T.~Ng, and M.~Zangeneh.
\newblock Dissipative particle dynamics: Introduction, methodology and complex
  fluid applications - a review.
\newblock {\em International Journal of Applied Mechanics}, 1\penalty0
  (04):\penalty0 737--763, 2009.

\bibitem{Mones2014}
L.~Mones, A.~Jones, A.~W. G{\"o}tz, T.~Laino, R.~C. Walker, B.~Leimkuhler,
  G.~Cs{\'a}nyi, and N.~Bernstein.
\newblock The adaptive buffered force {QM/MM} method in the {CP}\textup{2}{K}
  and {AMBER} software packages.
\newblock {\em Journal of Computational Chemistry}, 36\penalty0 (9):\penalty0
  633--648, 2015.

\bibitem{Nikunen2003}
P.~Nikunen, M.~Karttunen, and I.~Vattulainen.
\newblock How would you integrate the equations of motion in dissipative
  particle dynamics simulations?
\newblock {\em Computer Physics Communications}, 153\penalty0 (3):\penalty0
  407--423, 2003.

\bibitem{Nose1984a}
S.~Nos{\'e}.
\newblock A unified formulation of the constant temperature molecular dynamics
  methods.
\newblock {\em The Journal of Chemical Physics}, 81\penalty0 (1):\penalty0 511,
  1984.

\bibitem{Pan2014}
D.~Pan, N.~Phan-Thien, and B.~C. Khoo.
\newblock Dissipative particle dynamics simulation of droplet suspension in
  shear flow at low {C}apillary number.
\newblock {\em Journal of Non-Newtonian Fluid Mechanics}, 212:\penalty0 63--72,
  2014.

\bibitem{Pan2010}
W.~Pan, B.~Caswell, and G.~E. Karniadakis.
\newblock Rheology, microstructure and migration in {B}rownian colloidal
  suspensions.
\newblock {\em Langmuir}, 26\penalty0 (1):\penalty0 133--142, 2010.

\bibitem{Pastorino2007}
C.~Pastorino, T.~Kreer, M.~M{\"u}ller, and K.~Binder.
\newblock Comparison of dissipative particle dynamics and {L}angevin
  thermostats for out-of-equilibrium simulations of polymeric systems.
\newblock {\em Physical Review E}, 76\penalty0 (2):\penalty0 026706, 2007.

\bibitem{Peters2004}
E.~A. J.~F. Peters.
\newblock Elimination of time step effects in {DPD}.
\newblock {\em Europhysics Letters}, 66\penalty0 (3):\penalty0 311, 2004.

\bibitem{Pivkin2010}
I.~V. Pivkin, B.~Caswell, and G.~Karniadakis.
\newblock Dissipative particle dynamics.
\newblock {\em Reviews in Computational Chemistry}, 27:\penalty0 85--110, 2010.

\bibitem{Pivkin2005}
I.~V. Pivkin and G.~E. Karniadakis.
\newblock A new method to impose no-slip boundary conditions in dissipative
  particle dynamics.
\newblock {\em Journal of Computational Physics}, 207\penalty0 (1):\penalty0
  114--128, 2005.

\bibitem{Revenga1999}
M.~Revenga, I.~Z{\'u}{\~n}iga, and P.~Espa{\~n}ol.
\newblock Boundary conditions in dissipative particle dynamics.
\newblock {\em Computer Physics Communications}, 121-122:\penalty0 309--311,
  1999.

\bibitem{Rugh1997}
H.~H. Rugh.
\newblock Dynamical approach to temperature.
\newblock {\em Physical Review Letters}, 78\penalty0 (5):\penalty0 772, 1997.

\bibitem{Serrano2006}
M.~Serrano, G.~De~Fabritiis, P.~Espa{\~n}ol, and P.~V. Coveney.
\newblock A stochastic {T}rotter integration scheme for dissipative particle
  dynamics.
\newblock {\em Mathematics and Computers in Simulation}, 72\penalty0
  (2):\penalty0 190--194, 2006.

\bibitem{Shang2015}
X.~Shang, Z.~Zhu, B.~Leimkuhler, and A.~J. Storkey.
\newblock Covariance-controlled adaptive {L}angevin thermostat for large-scale
  {B}ayesian sampling.
\newblock In {\em Advances in Neural Information Processing Systems 28}, pages
  37--45, 2015.

\bibitem{Shardlow2003}
T.~Shardlow.
\newblock Splitting for dissipative particle dynamics.
\newblock {\em SIAM Journal on Scientific Computing}, 24\penalty0 (4):\penalty0
  1267--1282, 2003.

\bibitem{Shardlow2006}
T.~Shardlow and Y.~Yan.
\newblock Geometric ergodicity for dissipative particle dynamics.
\newblock {\em Stochastics and Dynamics}, 6\penalty0 (01):\penalty0 123--154,
  2006.

\bibitem{Sivak2013}
D.~A. Sivak, J.~D. Chodera, and G.~E. Crooks.
\newblock Using nonequilibrium fluctuation theorems to understand and correct
  errors in equilibrium and nonequilibrium simulations of discrete {L}angevin
  dynamics.
\newblock {\em Physical Review X}, 3\penalty0 (1):\penalty0 011007, 2013.

\bibitem{Soddemann2003}
T.~Soddemann, B.~D{\"u}nweg, and K.~Kremer.
\newblock Dissipative particle dynamics: A useful thermostat for equilibrium
  and nonequilibrium molecular dynamics simulations.
\newblock {\em Physical Review E}, 68\penalty0 (4):\penalty0 046702, 2003.

\bibitem{Stoyanov2005}
S.~D. Stoyanov and R.~D. Groot.
\newblock From molecular dynamics to hydrodynamics: A novel {G}alilean
  invariant thermostat.
\newblock {\em The Journal of Chemical Physics}, 122:\penalty0 114112, 2005.

\bibitem{Suter2014}
J.~L. Suter, D.~Groen, and P.~V. Coveney.
\newblock Chemically specific multiscale modeling of clay--polymer
  nanocomposites reveals intercalation dynamics, tactoid self-assembly and
  emergent materials properties.
\newblock {\em Advanced Materials}, 27\penalty0 (6):\penalty0 966--984, 2014.

\bibitem{Symeonidis2005}
V.~Symeonidis, G.~Karniadakis, and B.~Caswell.
\newblock A seamless approach to multiscale complex fluid simulation.
\newblock {\em Computing in Science and Engineering}, 7\penalty0 (3):\penalty0
  39--46, 2005.

\bibitem{Talay1990}
D.~Talay and L.~Tubaro.
\newblock Expansion of the global error for numerical schemes solving
  stochastic differential equations.
\newblock {\em Stochastic Analysis and Applications}, 8\penalty0 (4):\penalty0
  483--509, 1990.

\bibitem{Tang1995}
S.~Tang, G.~T. Evans, C.~P. Mason, and M.~P. Allen.
\newblock Shear viscosity for fluids of hard ellipsoids: {A} kinetic theory and
  molecular dynamics study.
\newblock {\em The Journal of Chemical Physics}, 102\penalty0 (9):\penalty0
  3794--3811, 1995.

\bibitem{Travis2008}
K.~P. Travis and C.~Braga.
\newblock Configurational temperature control for atomic and molecular systems.
\newblock {\em The Journal of Chemical Physics}, 128:\penalty0 014111, 2008.

\bibitem{Trozzi1984}
C.~Trozzi and G.~Ciccotti.
\newblock Stationary nonequilibrium states by molecular dynamics. {II}.
  {N}ewton's law.
\newblock {\em Physical Review A}, 29\penalty0 (2):\penalty0 916--925, 1984.

\bibitem{Vattulainen2002}
I.~Vattulainen, M.~Karttunen, G.~Besold, and J.~M. Polson.
\newblock Integration schemes for dissipative particle dynamics simulations:
  From softly interacting systems towards hybrid models.
\newblock {\em The Journal of Chemical Physics}, 116:\penalty0 3967, 2002.

\bibitem{Verlet1967}
L.~Verlet.
\newblock Computer ``experiments" on classical fluids. {I}. {T}hermodynamical
  properties of {L}ennard-{J}ones molecules.
\newblock {\em Physical Review}, 159\penalty0 (1):\penalty0 98, 1967.

\bibitem{Whittle2010}
M.~Whittle and K.~P. Travis.
\newblock Dynamic simulations of colloids by core-modified dissipative particle
  dynamics.
\newblock {\em The Journal of Chemical Physics}, 132\penalty0 (12):\penalty0
  124906, 2010.

\end{thebibliography}

\end{document}